\documentclass[aps,prb,twocolumn,amsmath,amssymb,superscriptaddress,floatfix,showpacs,preprintnumbers,reprint]{revtex4-2}

\usepackage{graphics,amssymb,amsmath,epsfig,color,textgreek}
\usepackage{graphicx}
\usepackage[dvipsnames]{xcolor}
\usepackage{dcolumn}
\usepackage{bm}
\usepackage[normalem]{ulem}
\usepackage[colorlinks=true,citecolor=cyan]{hyperref}
\hypersetup{colorlinks=true,citecolor=cyan,linkcolor=red,urlcolor=magenta}
\usepackage{braket}
\usepackage[capitalise]{cleveref}
\usepackage{lipsum}
\usepackage[dvipsnames]{xcolor}

\usepackage{cancel}

\usepackage[dvipsnames]{xcolor}

\newcommand{\ham}{\mathcal{H}}
\newcommand{\hammat}{\mathbb{H}}
\newcommand{\ess}{\mathbb{S}}
\newcommand{\tmax}{T_{\mathrm{max}}}

\newcommand{\taumax}{\tau_{\mathrm{max}}}
\newcommand{\Tmax}{T_{\mathrm{max}}}
\newcommand{\dErel}{\Delta \varepsilon_\mathrm{rel}}

\usepackage{comment}

\begin{document}

\title{Cheaper and more noise-resilient quantum state preparation using eigenvector continuation}

\author{Anjali A. Agrawal}
\affiliation{Department of Physics, North Carolina State University, Raleigh, North Carolina 27695, USA}

\author{Jo\~ao C. Getelina}
\affiliation{Department of Physics, North Carolina State University, Raleigh, North Carolina 27695, USA}

\author{Akhil Francis}
\affiliation{Department of Physics, North Carolina State University, Raleigh, North Carolina 27695, USA}
\affiliation{Applied Mathematics and Computational Research Division, Lawrence Berkeley National Laboratory, Berkeley, CA 94720, USA}

\author{A.~F.~Kemper}
\email{akemper@ncsu.edu}
\affiliation{Department of Physics, North Carolina State University, Raleigh, North Carolina 27695, USA}

\begin{abstract}

Subspace methods are powerful, noise-resilient methods that can effectively prepare ground states on quantum computers.
The challenge is to get a subspace with a small condition number that spans the states of interest using
minimal quantum resources.
In this work, we will use eigenvector continuation (EC) to build a subspace from the low-lying states of a set
of Hamiltonians.  The basis vectors are prepared using
truncated versions of
standard state preparation methods such as imaginary time evolution (ITE), adiabatic state preparation (ASP), and variational quantum eigensolver (VQE).
By using these truncated methods combined with eigenvector
continuation, we can
directly improve upon them, obtaining more accurate
ground state energies at a reduced cost.
We use several spin systems 
to demonstrate convergence 
even when methods like ITE and ASP fail, 
such as ASP in the presence of level crossings and ITE with vanishing energy gaps.
We also showcase the noise resilience of this approach
beyond the gains already made by having shallower quantum
circuits. Our findings suggest that eigenvector
continuation can be used to improve existing
state preparation methods in the near term.

\end{abstract}

\maketitle

\section{Introduction}

Finding the ground state of non-integrable systems
is one of the primary targets of studies in quantum
mechanics. 
One can achieve these goals via classical numerical methods, such as exact diagonalization or tensor network algorithms.
However, these methods have limitations, mainly for two- and three-dimensional systems and/or for highly entangled states~\cite{sandvik2010computational,ueda2023finite}. 
As a result, there has been significant interest in quantum computing alternatives to address this issue.
Within the setting of quantum computing,
the problem is recast as finding an efficient
way to prepare a desired quantum state.
This problem is generally QMA-hard \cite{kitaev2002classical, kempe2006complexity, chen2024local}, i.e.
difficult even for quantum computers; nevertheless,
a significant research effort is being directed
towards making inroads into this important problem,
as quantum devices are believed to be able to approach this problem and perform much better than their classical counterparts \cite{zalka1998simulating, daley2022practical}. 
This promise is not yet realized, and problems such
as barren optimization plateaus \cite{cerezo2021variational,ragone2023unified, ortiz2021entanglement}
plague the current methods. While fault tolerant
algorithms that have provable convergence rates
do exist, these are not yet realizable on near-term
hardware, which is noisy and limited in its
capabilities.
Because of this reason, researchers are focusing on developing NISQ-friendly algorithms that can still perform well with hardware imperfections \cite{preskill2018quantum, Huang2023near, oftelie2021simulating}.

Several quantum state preparation algorithms can be categorized into these broad categories, including variational quantum algorithms (e.g, VQE, QAOA) \cite{VQE,mcclean2016theory,omalley2016scalable,tilly2022review,liu2022stochastic, meitei2021gatefree, shang2023schrodingerheisenberg, akshay2022circuit, tubman2018postponing, benedetti2021hardware}, time evolution methods (e.g, ASP, QITE) \cite{farhi_quantum_2000,hejazi_adiabatic_2023, kovalsky2023self, tsuchimochi_improved_2023, vanvreumingen2023adiabatic, mc2024towards}, and subspace methods (e.g, Krylov subspace) \cite{asthana_quantum_2023, cortes2022fast, francis2020quantum, cortes2022quantum, Lanczos1952SolutionOS, zhang2023measurementefficient}. 
While they all have advantages and disadvantages, none are ideal for implementation on current hardware.
Variational quantum algorithms use classical optimization-based and learning-based approaches with parameterized quantum circuits but face optimization landscape issues along with trainability, accuracy, and efficiency limitations \cite{cerezo2021variational}.
Time evolution methods, and in particular adiabatic state preparation (ASP), provide an efficient mapping to a quantum circuit; 
however, ASP often requires a long evolution \cite{Childs2021theory,francis2022adiabatic,kovalsky2023self}, an easily prepared starting state, and cannot be used in the presence of symmetry-protected level crossings.
Imaginary time evolution \cite{motta2020determining} (ITE) guarantees convergence, but implementing non-unitary evolution in a quantum circuit is complicated. 
Both evolution-based algorithms tend to involve deep quantum circuit implementations;
and as the system size and the corresponding Hilbert space
grow, these algorithms
require a concomitantly longer circuit to converge
 \cite{Heras2014Digital, younis2021qfast, kokcu2022fixed, kokcu2022algebraic},
thus increasing the quantum resources required.

An alternate approach can be found within quantum subspace methods,
where the Hamiltonian is projected onto a smaller subspace, and after obtaining the relevant matrix elements the problem is recast as a generalized eigenvalue problem \cite{asthana_quantum_2023, cortes2022fast, francis2020quantum, cortes2022quantum, zhang2023measurementefficient, Epperly2022theory}.
Subspace methods do not require optimizing parameterized circuits,
avoiding issues of optimization landscapes like barren plateaus \cite{cerezo2021variational}.
The price to pay is to find a subspace that spans
the desired region of Hilbert space;
moreover,
the techniques developed for this purpose often face the issue of getting a desired subspace, 
i.e. one that has enough overlap with the ground state of the system of interest and avoids an ill-conditioned subspace overlap matrix.
For example, in methods that use a Krylov subspace, 
the basis states are generated by applying operators obtained from the same Hamiltonian, and this can result in quite similar basis states that have high overlap with each other \cite{Lanczos1952SolutionOS, calvetti_regularizing_2002, bharti2022noisy, getelina2024quantum},
and thus form an ill-conditioned problem.

One method that can avoid this issue is
eigenvector continuation \cite{Frame_2018,francis2022subspace,duguet2023eigenvector},
where the subspace basis vectors are generated from
different but related Hamiltonians. 
When a parameter of a system Hamiltonian is varied smoothly, the set of the ground states is typically spanned by a few low-energy vectors. This statement implies that we can get a subspace by obtaining the ground state---using a method of choice---at a few ``training points'' 
in the parameter range of interest and cheaply finding ground states at all the remaining parameter points by solving a small generalized eigenvalue problem (GEP) at each point (see \cref{fig:algo_fig}).
This method has been used in multiple fields such as quantum chemistry and nuclear physics \cite{yapa_volume_2022, konig_eigenvector_2020, Frame_2018, Mejuto_Zaera_2023,duguet2023eigenvector}.

In this work, we will improve on this idea by realizing that an exact ground state is not needed for EC to work; rather, obtaining a number of states with a decent overlap
with the low energy manifold suffices.
Here we will use real- and imaginary-time ground-state preparation techniques, although other methods could be used similarly to get the low-energy states.
In all cases, an otherwise insufficiently deep (or truncated) circuit can be used.
Getting low-energy states is much easier and more cost-effective than obtaining exact ground states; we will explore these ideas for time evolution methods in the next section.
We will show how this method can be used to handle issues with time-evolution methods like vanishing energy gaps and/or level crossings,
and saving resources in doing so.
A similar idea of using the approximate state preparation for getting a basis for subspace diagonalization was explored in a recent work \cite{hirsbrunner2024diagnosing}, where they use approximate VQE for preparing a subspace to solve quantum chemistry instances, but remain focused on a single Hamiltonian.

The outline of the paper is as follows: In section II, we will outline the method of eigenvector continuation.
In section III, we will apply it using time evolution and variational state preparation methods to get a low-energy subspace for the XY spin model. 
In section IV, we will extend the application of the method to address the issue of vanishing energy gaps using the time-evolution methods with an example of a model with level crossings, XY model, and model with changing degeneracies in the ground state, Kagome XXZ model.  
Section V discusses the noise resilience of the method by performing noisy simulations. 
Lastly, we conclude with a discussion in section VI.

\section{Eigenvector Continuation}

Eigenvector continuation (EC) has been used previously in many areas showing promising results with better convergences and optimized use of resources \cite{yapa_volume_2022, konig_eigenvector_2020, Frame_2018, Mejuto_Zaera_2023, francis2022subspace}. 
This work is based on the same formalism of eigenvector continuation (see appendix \ref{app:EC} for further details),
which can be summarized as follows --- 
for a set of Hamiltonians indexed by a parameter $\ham_{target}(\theta)$, the ground states are spanned by a few low-energy vectors of the Hamiltonians taken at only a few parameter points.
In other words, if we consider two Hamiltonians that are close, their low-energy state profile is related and those states will have high overlap with each other.
We can exploit this feature to obtain a low-energy subspace from a few ($k_p$) training points 
and get a ground state energy spectrum for the full set of target Hamiltonians $\ham_{target}(\theta)$.

For EC to work, we require enough basis states that are well spread in the varying parameter space to span the whole low energy subspace.
The precise details of how many basis states and how to choose the training points
highly depends on the properties of the Hamiltonian  
such as the presence of underlying symmetries, level crossings in the energy spectrum, and degeneracy of the ground state among others. 
This is discussed in more detail in the next sections for specific systems. 

In this work, we still use a set of basis states that spans the low-energy subspace, but we propose using state preparation methods that are truncated before they reach the ground state. 
This gives us a low-energy state that is not necessarily an eigenstate or the ground state, rather, this is a superposition of a few low-energy states of the system. 
As we will show, getting these states from the small number of training points gets us enough overlap with all the ground states to be able to obtain reasonable ground states from the generalized eigenvalue problem.

\begin{figure}[htpb]
    \centering
    \includegraphics[clip= true, trim=1.4cm 4.3cm 16.7cm 0cm,width=0.48\textwidth]{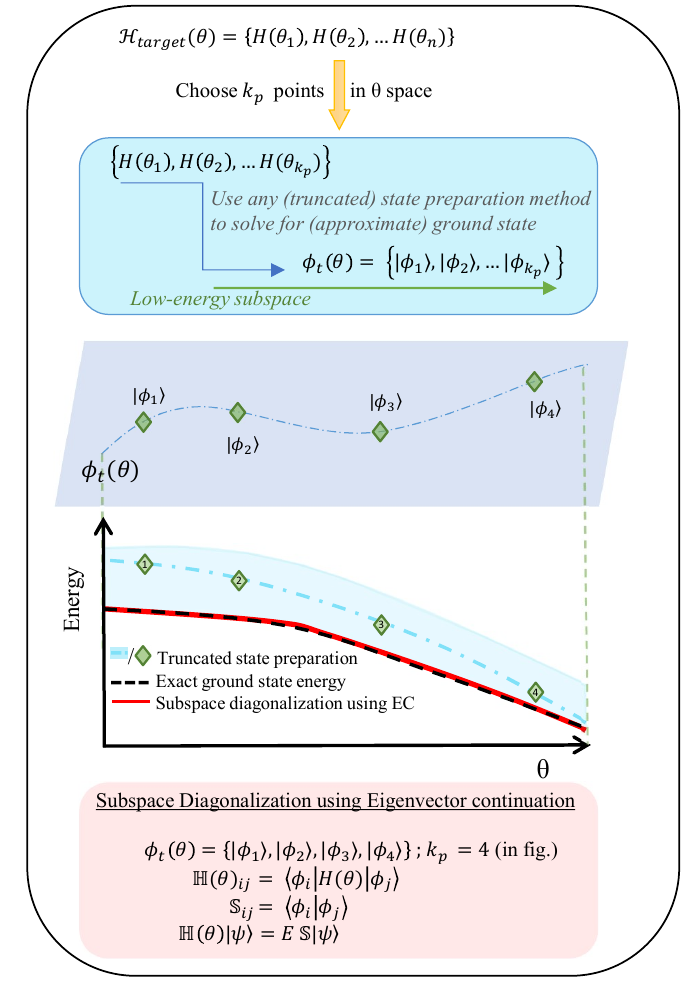}
    \caption{Layout of the method -- EC with Truncated state preparation. Truncated state preparation generated approximate ground state in vector space $\phi_t(\theta)$ for training point Hamiltonian set, chosen $k_p$ points from the target set $\ham_{target}$. This basis is then used for subspace diagonalization for the full set of $\ham_{target}$.
    }
    \label{fig:algo_fig}
\end{figure}

The method is sketched out in \cref{fig:algo_fig}. 
Given a Hamiltonian $\mathcal{H}$ that varies with a (possibly multidimensional) parameter $(\theta)$, the task is to find the low energy spectrum of a set of Hamiltonians
$\{H(\theta_1),H(\theta_2),\ldots,H(\theta_n)\}$.
From the set,  Hamiltonians at $k_p$ different points in the $\theta$ space are chosen as training points; 
the choice can incorporate any prior knowledge about the system such as the presence of conserved quantities. 
In any case, the points should be chosen in such a way
that
the full space of interest is covered,
especially in cases where the ground state changes due to the presence of protected crossings \cite{herbst2022surrogate}.
Having training points very close to each other can produce a desirable subspace in certain cases, however,
this can also cause ill-conditioning of the overlap matrix from over-spanning the space as very close parameter-spaced basis states might produce the states whose overlap is near unity \cite{Epperly2022theory}.

For the training Hamiltonians, we find {\emph{approximate}} ground states, using any state preparation method, to obtain a set of low-energy vectors that will form the basis vectors for a subspace $(\phi_t(\theta) = \{ \ket{\phi_1}, \ket{\phi_2}, \dots \ket{\phi_{k_p}} \} )$. 
In this paper we have focused on time evolution methods for state preparation---ASP and ITE; but we also provide one example using a variational method to showcase the versatility of our approach.

Now that we have a subspace that spans low energy states, we project the target Hamiltonians $(H(\theta_i))$ onto this subspace and solve the generalized eigenvalue problem: 
\begin{align}\label{eq:GEP}
    \hammat_{ij}(\theta) &= \braket{\phi_i|H(\theta)|\phi_j} \nonumber\\
    \ess_{ij} &= \braket{\phi_i|\phi_j}  \\
    \hammat(\theta) \ket{\psi} &= E \ess \ket{\psi}.\nonumber
\end{align}
One can solve Eq.~\eqref{eq:GEP} via standard eigendecomposition routines. 
Ill-conditioned cases occur whenever $\ess$  is non-positive definite, thus requiring a separate diagonalization of $\ess$, 
followed by applying some truncation scheme for the corresponding eigenpairs and a basis rotation of $\hammat$~\cite{Epperly2022theory, getelina2024quantum}. 
These extra steps were unnecessary in this work, as we have a positive-definite overlap matrix for all the cases studied here.

The lowest eigenvalue thus obtained gives the ground state energy of the target Hamiltonian.
The overlap matrix $(\ess)$ of the basis vectors remains the same for all the target points, but the projected Hamiltonian ($\hammat$) needs to be calculated separately for each target point.
We can use the fact that the target Hamiltonians differ only by the values of the coefficients of the Hamiltonian operators 
and simplify the process of measuring the projected Hamiltonians ($\hammat$), measuring all the Hamiltonian operators separately and adding them with corresponding coefficients for each target point.

\section{Improving Quantum state preparation methods using EC}

In this work, we will use time evolution-based methods, 
i.e. adiabatic state preparation (ASP) and imaginary time evolution (ITE), for generating our low-energy subspace.
For completeness, we also show one example using the variational quantum eigensolver (VQE) to obtain the training points. 
However, contrary to time evolution methods, there is no straightforward way to truncate VQE; therefore, we dedicate most of our presentation to results obtained via ASP and ITE.

To improve the quality of the state prepared via time evolution methods, one can consider either a longer evolution time, or a smaller time step; we will focus on the former.
To illustrate the use of EC to improve a truncated state preparation, we truncate
at $\leq 10\%$ of the time evolution steps required to get reasonable ground state energies. 

In what follows, we briefly describe the state preparation methods considered here.

\subsubsection{Imaginary Time Evolution}

One of the methods that we will use to get a low-energy subspace is truncated imaginary time evolution.
Imaginary time evolution \cite{motta2020quantum} guarantees convergence to the ground state of a Hamiltonian $\ham$ as long as we start with a state $\ket{\psi}$ with non-zero overlap with the ground state of the system.
We evolve such a state in imaginary time via
\begin{align}\label{eq:ITE}
    \ket{\psi(\tau)} = e^{-\tau \ham} \ket{\psi(\tau=0)},
\end{align}
to some maximum imaginary time $\taumax$,
normalizing at every time step.
As we evolve, the overlap with the higher energy states vanishes rapidly, whereas states close to the ground state take longer evolution times (see Appendix \ref{app:QITE}). 
Since the goal is to obtain a subspace with low-energy vectors, we need not eliminate all of these low-energy states. 
In fact, they will be useful for spanning the space with ground states across the parameter range, 
since the first few excited states for one Hamiltonian at one parameter point are potential ground states at others.
This lets us stop the evolution early, saving computational costs. 
We will discuss the overall resource comparison later in the context of the example model.

\subsubsection{Adiabatic State Preparation}
\label{sec:methods_ASP}

The other method we use is truncated real-time evolution or adiabatic state preparation.
In adiabatic state preparation, we start with a ground state of the system at a parameter value that is easier to prepare $ \ket{\psi_{gs}(t=0)}$ and adiabatically evolve in real-time to get the ground state at the target parameter valued Hamiltonian $\ket{\psi_{gs}(\mathcal{H}_N)}$. 
The time evolution is performed by
a unitary operator $U_N(t)$ that evolves the ground state at time $(t=0)$ to final time $(t=\Tmax)$ in $N$ time steps:
\begin{align} \label{eq:ASP}
    \ket{\psi_{gs}(\ham(t_N))} & = U_N (t) \ket{\psi_{gs}(\ham(t=0))} \nonumber  \\ 
    U_N(t) &\approx e^{-i dt \ham_N} \dots e^{-i dt \ham_2} e^{-i dt \ham_1}  + \mathcal{O}(dt^2)
\end{align}
where $dt=\frac{T}{N}$ and $\ham_j$ is the Hamiltonian at the $j$-th time step.
The error in approximating this unitary operator with first-order Trotter discretization is $\mathcal{O}(dt^2)$, which demands a large $N$.  

Since our Hamiltonian terms need not commute, 
we will use the Trotter-Suzuki decomposition \cite{trotter1959on, suzuki1976generalized, Childs2021theory} to approximate each evolution operator of the form $e^{-i dt \ham}$ given in Eq.~\ref{eq:ASP}.
For all the calculations in this manuscript, we will use trotterized gates for both ASP and ITE. 

This method works well if the evolution is slow --- i.e. if $d\ham/dt$ is smaller than the squared minimum gap $\Delta^2$ \cite{vanvreumingen2023adiabatic} --- and there are no protected level crossings between the ground state and excited states. When implementing ASP, one has a choice of protocol in how the Hamiltonian is varied; here we use a linear ramp,
and thus $d\ham/dt$ is set directly by $\Tmax$.
If $\Tmax$ is too small, i.e. if the Hamiltonian changes too quickly, the 
state undergoes diabatic excitation into the excited states, 
thus reducing the overlap with the
ground state and increasing the overlap with the
excited states (see Appendix \ref{app:ASP}). 
Again, we can use this to our advantage and form a subspace out of
non-adiabatic evolution using a small $\Tmax$, which can be performed
more cheaply than adiabatic evolution.

\subsubsection{Variational Quantum Eigensolver}
\label{sec:methods_VQE}

The Variational Quantum Eigensolver (VQE) is one of the most promising methods for ground-state preparation in the NISQ era. 
It stems from the variational principal of quantum mechanics, which states that the expectation value of a Hamiltonian $\ham$ with respect to a generic parameterized state $\left| \psi\left(\boldsymbol{\theta}\right) \right\rangle$ is bounded from below by the Hamiltonian ground-state energy $E_0$, i.e.,
\begin{equation}
    E_0 \le \left\langle \psi\left(\boldsymbol{\theta}\right) \right| \ham \left| \psi\left(\boldsymbol{\theta}\right) \right\rangle
\end{equation}
Therefore, by defining a parameterized ansatz that builds $\left| \psi\left(\boldsymbol{\theta}\right) \right\rangle$, one can construct the corresponding quantum circuit and measure the energy, 
which is fed into a classical minimizer routine as the cost function. The minimizer then returns a new set of optimized parameters $\boldsymbol{\theta}$. 
One repeats this procedure until the classical minimizer is converged, thus yielding an approximated ground state of the underlying system described by $\ham$.

In general VQE applications, one builds the state $\left| \psi\left(\boldsymbol{\theta}\right) \right\rangle$ by applying many layers of a predefined unitary operator $U\left(\boldsymbol{\theta}\right)$ to a classical product state $\left|\psi_0\right\rangle$, i.e.,
\begin{equation}
    \left| \psi\left(\boldsymbol{\theta}\right) \right\rangle = \prod_{\ell} U\left(\boldsymbol{\theta}_\ell\right) \left|\psi_0\right\rangle
\end{equation}
However, selecting the optimal ansatz $U\left(\boldsymbol{\theta}\right)$ is a problem-dependent question that has led to many different proposals of VQE ansatze~\cite{tilly2022review}. 
In this work we choose the one known as Hamiltonian Variational Ansatz (HVA), which, as the name suggests, takes the operators that appear in the system Hamiltonian to build the variational state~\cite{Wecker2015VHA,wiersema2020exploring}. 
We explicitly show the ansatz considered here in the next section, after the system Hamiltonian is introduced.

In addition, differently from ASP and ITE, VQE does not have a single variable to be chosen as truncation parameter; 
for instance, one could choose to truncate the number of variational parameters $\boldsymbol{\theta}$, 
or the number of ansatz layers, or the number of iterations within the classical optimizer. 
In this work we choose the latter as our truncation parameter. We also use the maximum number of variational parameters (i.e., one per gate) and keep the number of HVA layers as low as possible.

\subsection{Application to 5-site XY spin chain}

We demonstrate these approaches for a test
model: a 5-site XY open spin chain with mixed magnetic fields,

\begin{align} \label{XYham}
    \ham =& J\sum_{i=1}^{N-1} \bigg[ X_i X_{i+1} + Y_i Y_{i+1} \bigg] \nonumber \\
    &+ \sum_{i=1}^N \bigg[B_Z Z_i + (-1)^i B_X X_i \bigg]
\end{align}

where we fix $N=5$, $J=1$ and a staggered $B_X = 0.2$, and vary 
$B_Z \in [0,3]$. 
A finite staggered transverse field in the $X$ direction ($B_X$) opens up an energy gap at level crossings --- this is necessary to lift
the degeneracy at level crossings, allowing adiabatic time evolution to follow the lowest energy state (see \cref{fig:XYspectrum}). 
In the next section, we will address these issues in detail.

\begin{figure}
    \centering
    \includegraphics[clip= true, trim=6.2cm 5.2cm 15.2cm 10cm, width=0.38\textwidth]{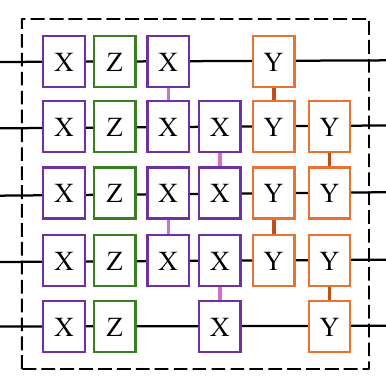}
    \caption{Schematic of a single layer of the Hamiltonian Variational Ansatz (HVA) for a 5-site XY model with transverse and longitudinal external fields.}
    \label{fig:HVA-circuit}
\end{figure}

Hence, given the Hamiltonian in Eq.~\eqref{XYham}, one layer of the HVA ansatz corresponds to 
\begin{equation}
    U\left(\boldsymbol{\theta}\right)=U_{YY}\left(\boldsymbol{\delta}\right)U_{XX}\left(\boldsymbol{\gamma}\right)U_{Z}\left(\boldsymbol{\beta}\right)U_{X}\left(\boldsymbol{\alpha}\right)\label{eq:HVA-layer}
\end{equation}
where
\begin{align}
    U_{YY}\left(\boldsymbol{\delta}\right) & = \textrm{exp}\left(-i\sum_{\left\langle i,j\right\rangle}\delta_{i,j}Y_iY_j\right) \nonumber\\
    U_{XX}\left(\boldsymbol{\gamma}\right) & = \textrm{exp}\left(-i\sum_{\left\langle i,j\right\rangle}\gamma_{i,j}X_iX_j\right) \nonumber\\
    U_{Z}\left(\boldsymbol{\beta}\right) & = \textrm{exp}\left(-i\sum_i\beta_iZ_i\right) \nonumber\\
    U_{X}\left(\boldsymbol{\alpha}\right) & = \textrm{exp}\left(-i\sum_i\alpha_iX_i\right) \nonumber
\end{align}
Fig.~\ref{fig:HVA-circuit} depicts the quantum circuit representation of Eq.~\eqref{eq:HVA-layer}. 
For the simulations of the 5-site XY model, we have considered an ansatz with two HVA layers, with each layer following the same operator ordering as shown in Fig.~\ref{fig:HVA-circuit}.

For our analysis, we will consider $20$ equally distributed points in varying $B_Z$ range as target Hamiltonians.
For those points, we calculate the ground state energy eigenvalues solving the GEP (\cref{eq:GEP}) with the subspace formed by truncated states. 
To compare errors with different levels of truncation, 
we will consider RMS error for all the target points, given by:
\begin{equation}\label{eq:RMS}
    \Delta\varepsilon_{rms} = \sqrt{ \frac{1}{p}\sum_{i = 1}^{p} \bigg({|E_{cal}^i-E_{exact}^i|} \bigg)^2 }
\end{equation}

\begin{figure*}[htpb]
    \centering
    \includegraphics[width=0.94\textwidth]{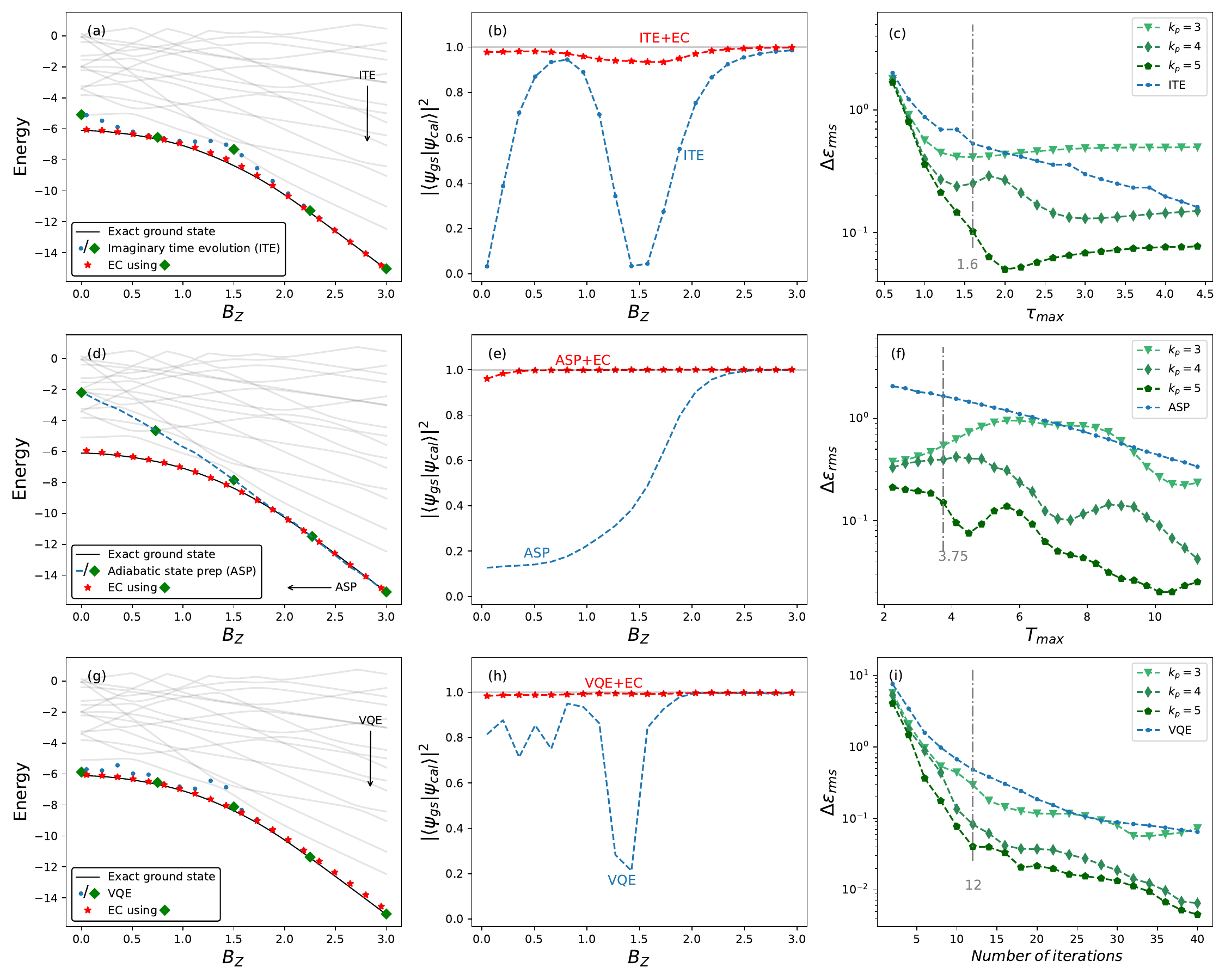}
    \caption{Implementing EC with truncated ASP, ITE, and VQE for a 1D chain XY model 5 site. Panels (a), (d), and (g) show the system complete energy spectrum (full lines) as a function of the longitudinal field $B_z$, with the ground-state energy curve highlighted in black. The green diamonds represent the reference points we used for building the subspace and computing the corresponding EC curve (red stars). The arrows indicate the direction taken by each state preparation method. Panels (b), (e), and (h) compare the fidelities of EC and truncated state preparation data with respect to the exact ground state. Panels (d), (f), and (i) show the energy error [Eq.~\eqref{eq:RMS}] as a function of truncation parameters, which are evolution time for ITE and ASP and number of iterations in the minimizer routine for VQE. Here we have considered three different subspace dimensions $k_p$ for obtaining the EC data, which we also compare to the truncated state preparation results without EC (blue dashed curve). The vertical lines represent the truncation parameter value used for the data shown in the other panels.}
    \label{fig:group_XY5}
\end{figure*}

In \cref{fig:group_XY5} we demonstrate how using eigenvector continuation improves the
ground state preparation algorithms.
For the time evolution methods, we considered a modest number of time steps (40 and 75 for ITE and ASP, respectively) to get the subspace basis vectors.
For comparison, obtaining a reasonable ground state 
would require 600 and 750 steps for ITE and ASP respectively, for all of the target points. 
For the truncated VQE, we considered 12 iterations of the classical optimizer for all the training points. For comparison, the unbounded minimizer would require from 73 to 869 iterations to converge for these exact points.
To get the basis vectors, we choose equally spaced $k_p=5$ training points for the energy spectrum and fidelity analysis and $k_p = \{3,4,5\}$ training points for the point of truncation analysis.    

First, let us discuss the ITE results.
We use $d\tau = 0.2$ and $\taumax = 1.6$, i.e. $8$ evolution steps for each training point Hamiltonian.
Panel (a) shows the spectrum of the model, together with
the final energies from the truncated ITE, and the improved results obtained
when EC is used on top of the truncated ITE results.
As expected, truncated ITE (blue dots) does not achieve the ground
state energy for all parameter values --- in the few
cases where it does, the ground state is well-separated
from the excited states.  
We take $k_p = 5$ equally-spaced
truncated ITE results (green diamonds), use these as a subspace
basis for EC and compute the spectrum. 
The EC spectrum (red stars) matches the exact solution considerably better
than the truncated ITE by itself.  
This is corroborated
by the fidelity, shown in panel (b). The fidelity of truncated
ITE is near zero in the worst cases and varies significantly
throughout the spectrum.  Using EC significantly improves the situation.

Moving to truncated ASP, we again show the spectrum together with the
obtained energies from truncated ASP in panel (d).
Here, we have used $\Tmax = 3.75$, and $dt=0.05$, i.e.
75-time steps, and we use a linear ramp in $B_z$ with $dB_z/dt = 0.8$.
The process starts at $B_Z=3$, where the ground state
is close to the fully polarized state, and thus is easy to prepare. 
The protected level crossings are turned into avoided crossings by the
$B_X=0.2$ which induces gaps far smaller than would be
appropriate for this particular ramp. 
Not surprisingly, the truncated ASP does not manage to follow
the ground state due to early truncation of the evolution, 
as the Hamiltonian is changing too rapidly for the adiabatic theorem to hold.
Nevertheless,
we can use the truncated ASP states to produce a
subspace, in which the ground state is accurately
captured. 
The resulting EC spectrum and fidelity
are shown in panels (d) and (e), respectively, where we
can observe that EC produces a marked improvement in both quantities.

Finally, panels (g) and (h) show the energies and fidelities calculated from the truncated VQE subspace. 
In this case, the selected training points (i.e., green diamonds) are closer to the exact ground state curve than the other two methods. 
Nevertheless, similarly to ITE and ASP, we observe a considerable improvement in the fidelities after we perform the subspace expansion, with the minimum fidelity for the considered $B_z$ range jumping from 21\% (blue dashed curve) to 98\% (red stars); a significant gain with little computational cost.

To make a quantitative comparison, we consider the RMS error ($\Delta\varepsilon_{rms}$) between the exact spectrum and the EC spectrum
for the 20 equally spaced target points under consideration (Eq.~\ref{eq:RMS}) 
using both methods.
We consider the RMS error as a function of $\taumax$ for ITE, which accounts for the total evolution time,  $\Tmax$ for
ASP, which accounts for the rate of change of the Hamiltonian, and the number of iterations of the classical optimizer for VQE. 
For all these methods, one can notice that the use of EC significantly improves the approximate results. 
At 5 training
points, which corresponds to the spectra and fidelities shown in \cref{fig:group_XY5}, we find a 78\% and 97\% reduction in error for ITE and ASP, respectively. 
The RMS error generally continues to 
reduce as
a function of $\taumax$/$\Tmax$. 
One notable exception occurs when $k_p = 3$,
where the error increases for some values of $\tau_\mathrm{max}$/$T_\mathrm{max}$. 
This illustrates the
use of choosing training points that do not lie in
parameter regions that are difficult to calculate ---
with 3 training points, one of the critical parameters
is right at an avoided level crossing, which leads
to the increase in the RMS error.
This issue can readily be solved by adding more training states.

For a comparison of resources required, let us first consider ASP.
To obtain an accurate ground energy spectrum for a set of target Hamiltonians, using ASP we require very slow (adiabatic) evolution. 
In our example, we would require $750$ time steps to get the fidelity close to $1$ (see \cref{fig:ATE_evolution}).
Using truncated ASP together with EC, using only $75$ time steps to get the basis vectors is sufficient (c.f. \cref{fig:group_XY5} (f)). 
Similarly, using ITE, the total time of evolution required for the desired error is $\taumax = 6$, i.e. $30$ imaginary time steps at each target point, yielding $30 \times 20 = 600$ total steps. 
With truncation, we used $\taumax=1.6$ which is $8$ time steps for $k_p = 5$ points, requiring $40$ total steps to get the basis vectors.
After getting the projected Hamiltonian and overlap matrix, all the energy values are calculated by solving GEP, 
which is classically easy given matrix sizes in GEP depending on the number of basis vectors. 
Overall, using EC on top of these standard state preparation techniques saves $>90 \%$ of the time evolution steps.

\subsection{Scaling with system size}

Although it is highly dependent on the nature of the system, the number of basis vectors required tends to increase polynomially with system size. 
In cases like the XY model, which exhibits different symmetry sectors in its ground states, 
the number of protected crossings is proportional to the number of lattice sites (see \cref{fig:XY8}),
and consequently, the subspace requires a similarly scaling number of basis states. 
For other systems like XXZ model (see appendix \ref{app:XXZ model}), where the ground state is doubly degenerate and has no protected crossings in the ground state, the growth is still polynomial but slower. 
We demonstrate this numerically using truncated ASP and EC
by calculating the RMS error as we increase the number of subspace
basis vectors $k_p$ for different system sizes of the XY and XXZ models.
We use the same parameters for the XY model as those in \cref{fig:group_XY5};
for the XXZ model, we set $J=+1$, $B_Z=0.2$ and vary $J_Z \in [0,3]$ using truncated ASP with parameters $dt=0.1$ and $\tmax = 1.2$.
Note that for the XXZ model ASP is easier to implement due to the absence of protected crossings;
the $B_Z$ field is introduced to break the (two-fold) degeneracy in the ground state.

Because the ground state energies vary as we change the number of sites, we cannot use $\Delta\varepsilon_{rms}$ directly. 
Instead, we will use the relative RMS error, defined as:
\begin{equation}\label{eq:relRMS}
    (\Delta\varepsilon_{rel})_{rms} = \sqrt{ \frac{1}{p}\sum_{i = 1}^{p} \bigg(\frac{|E_{cal}^i-E_{exact}^i|} {E_{exact}^i}\bigg)^2 }
\end{equation}
where the sum is over the $p=20$ target points,
$E_{exact}$ are exact ground state energies calculated using exact diagonalization, and $E_{cal}$ are the energies obtained using truncated ASP followed by EC. 

The low-energy subspace typically only contains a finite number of unique states.  As we increase the number
of training points, it is possible that we overspan the low-energy space, making the $\ess$ matrix non-invertible.
Thus, we only calculate errors for $k_p \leq (N_{XY}+2)$ for the XY model, and $k_p \leq N_{XXZ}$ for the XXZ model,
which was empirically determined based on when the subspace basis vectors become degenerate.

\begin{figure}[t]
    \centering
    \includegraphics[clip= true, trim=0cm 0.5cm 0cm 2.1cm,width=0.45\textwidth]{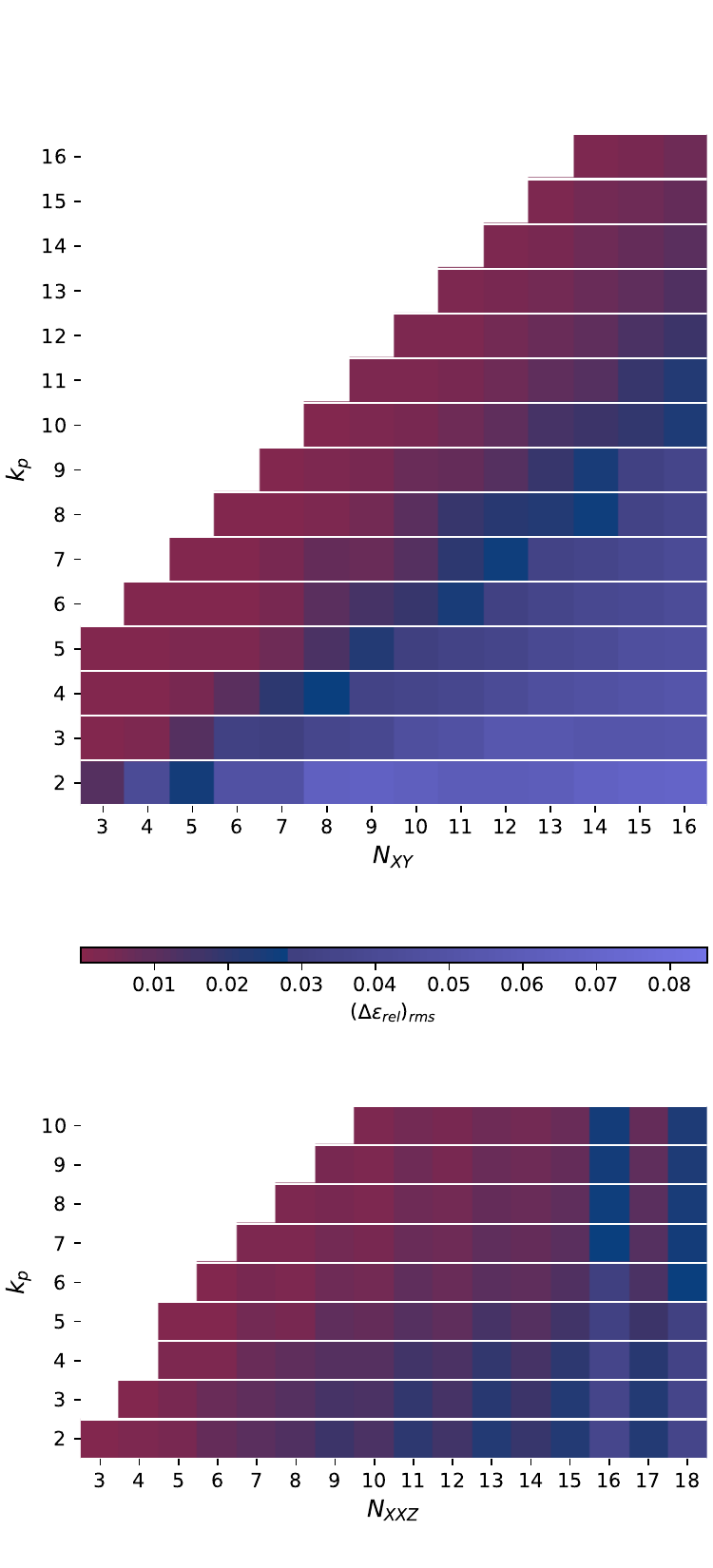}
    \caption{Relative RMS error in ground state energy eigenvalues for $20$ equally spaced points in parameter range $B_Z \in [0,3]$ for the different number of sites of XY model 1D chain and XXZ model 1D chain calculated using (truncated $ASP+EC$) with different number of training states ($k_p$) on the y-axis. White squares are not computed to address the over-spanning issue. 
    }
    \label{fig:NvsKp}
\end{figure}

The results are shown in \cref{fig:NvsKp}.  There is a general trend towards a decrease in the relative error as
the number of subspace vectors increases.  Because the low-energy subspace of the XY model
has more unique states than that of the XXZ model, the errors are larger, in particular when
$k_p$ is small compared to the number of sites $N_{XY}$.  The XXZ model, on the other hand,
has a relatively simple low-energy subspace and thus is spanned even with a small $k_p$. 
There is an odd-even pattern visible in the errors for the XXZ model that becomes more visible
at larger $N_{XXZ}$ that arises from details in the spectrum for this model --- we discuss this in \cref{app:XXZ model}.

For both of these models, we observe that the number of basis vectors required
for $(\dErel)_{rms} < 5\%$ scales linearly with the system size $N$, in stark contrast to the exponential scaling of the
Hilbert space dimension. 
For the XY model, this is because the number of phase transitions (or level crossings) in the ground energy spectrum is proportional to $N$ (see \cref{app:XY model}).
For the XXZ model, since the ground state changes smoothly without any level crossings (\cref{fig:ener_XXZ}), we required an even smaller number of training states, that again appear to scale linearly with system size ($N$).

\section{Avoiding protected level crossings and vanishing energy gaps}

One of the benefits of using EC as a subspace method is that the subspace basis can be generated using Hamiltonians
at points in parameter space that are easier to handle than the desired target parameters where state preparation
methods perform poorly.
For instance, ASP fails when the spectrum exhibits protected level crossings. In the previous section, these were
broken by a small $B_X$ field; however, it may be that the $B_X=0$ problem is of interest.
ITE can also have problems, and in particular, has issues with small energy gaps between the ground and excited
states. In this section, we will illustrate how to ameliorate this issue using EC.

\subsection{Getting around protected level crossings in ASP}

\begin{figure}[b]
    \centering
    \includegraphics[width=0.48\textwidth]{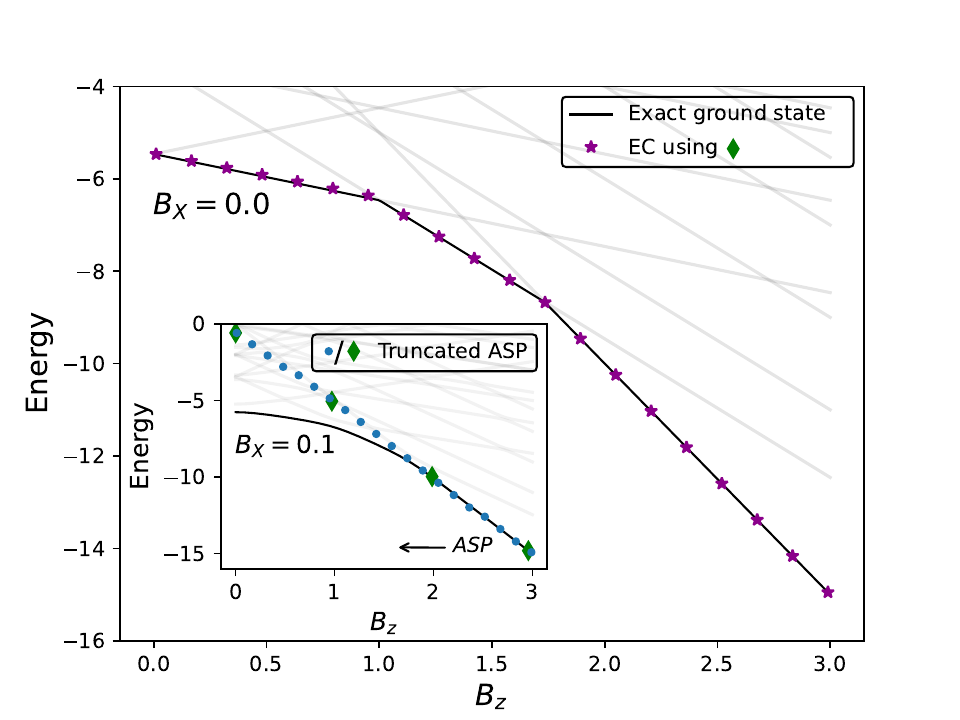}
    \caption{Ground energy spectrum for XY model 5-site with parameters $\{J=1, B_X=0.0\}$ calculated using EC with subspace obtained from ASP in parameter space with $\{J=1, B_X=0.1$\}}
    \label{fig:ATEBx0}
\end{figure}

As discussed in Sec.~\ref{sec:methods_ASP},
using ASP to generate a ground state requires finding a gapped adiabatic path, that is without any protected crossings \cite{farhi_quantum_2000}, which precludes its use for systems that exhibit level crossings.
The 1D transverse field XY model is such a system: in the absence of an in-plane magnetic
field ---  the spectrum shows several level crossings equal to the number of sites --- and thus
ASP is not an appropriate tool for finding the ground state.
However, we can get around this issue by performing ASP for the XY model that does have a small in-plane field,
and build a subspace using training states from this model.

In the \cref{fig:ATEBx0}, we perform ASP for a system with staggered $B_X=0.1$, $dt=0.05$ and $\tmax=3.75$, and choose four equally spaced states to form a basis. 
Next, we use this basis to find the eigenstates
of the target system with $B_X=0.0$.  
As is clear from the figure, this procedure readily
produces the correct ground state energies for the system of interest.

\subsection{Avoiding vanishing energy gaps in ITE}

\begin{figure}[htpb]
    \centering
    \includegraphics[width=0.48\textwidth]{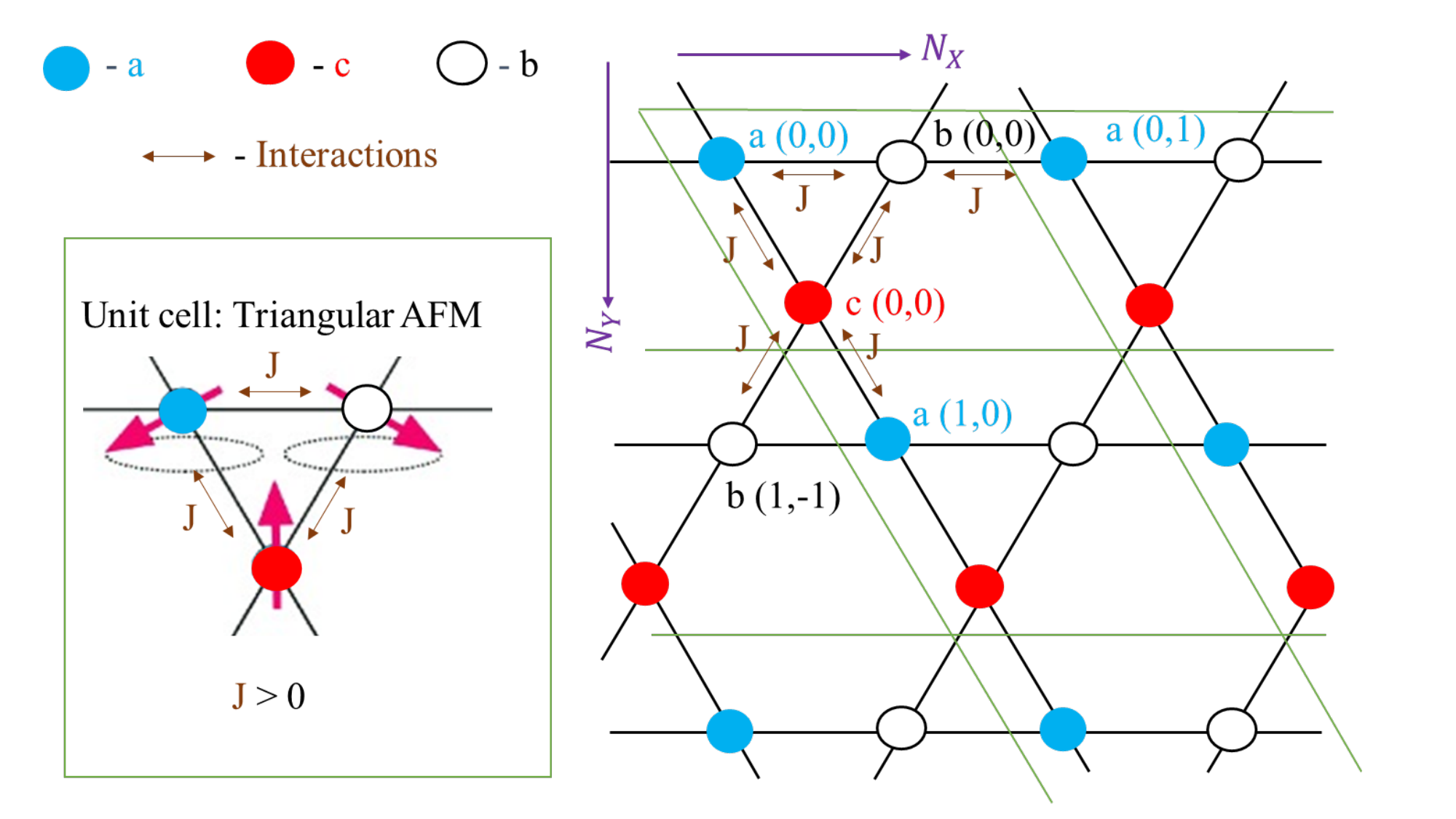}
    \caption{A 2D Kagome lattice with interaction strength $J>0$. A positive valued $J$ has an anti-ferro magnetic nature with frustration in unit cells shown in the green box. 
    $N_X$ and $N_Y$ are the number of unit cells in $x$ and $y$ axis respectively contained in the green lines. }
    \label{fig:Kagome_lattice}
\end{figure}

\begin{figure*}
    \centering
    \includegraphics[clip= true, trim=0cm 0cm 0.15cm 0cm, width=1.0\textwidth]{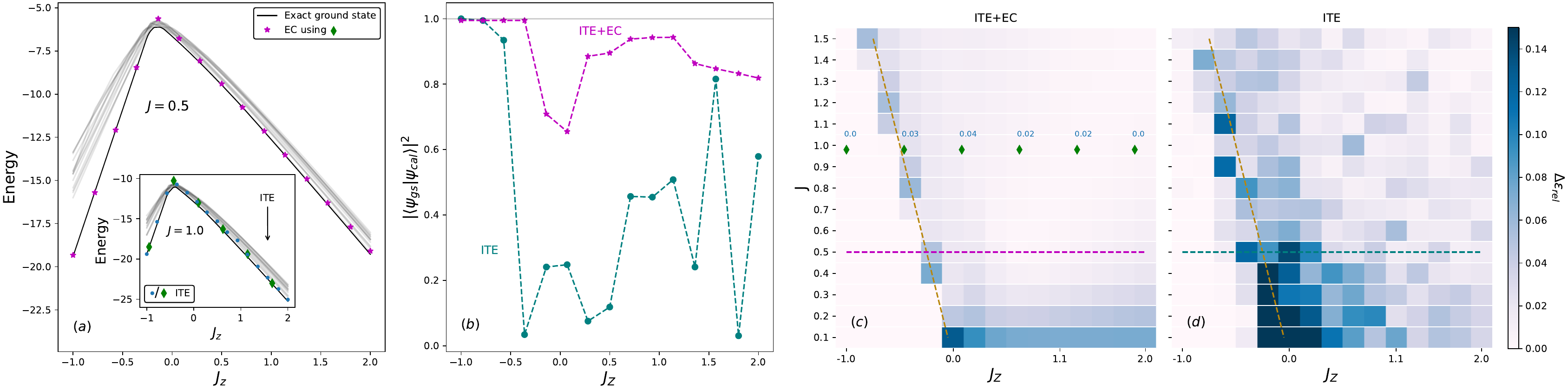}
    \caption{(a) Ground energy spectrum for Kagome 12 site XXZ model with parameters $J=0.5$ and $B_Z=0.2$ solved using subspace obtained from truncated ITE for $J=1.0$ and $B_Z=0.2$ (inset). 
    b) Fidelity of ITE and ITE+EC for $J=0.5$.
    c) $\Delta\varepsilon_{rel}$ for each target parameter point calculated using the same basis marked with green diamonds. The values marked on the diamonds are the relative errors in the basis vectors. 
    d) $\Delta\varepsilon_{rel}$ using just  ITE at all the parameter points. 
    In c) and d) The brown line indicates the phase transition line along ($J_Z/J = -0.5$).}
    \label{fig:group_kagome}
\end{figure*}

The idea of using a different Hamiltonian to generate a basis for a problem of interest
can be extended to address the issues in using ITE for systems with small and/or vanishing energy gaps.
To illustrate this, we will consider a more complex system: the 2D Kagome XXZ lattice.
Getting the ground energy spectrum for these systems of Hamiltonians is a difficult task due to the presence
of phase transitions, changing degeneracies in the ground state, and small energy gaps between the ground and
excited states \cite{Gotze2015ground}.
Consider a 2D Kagome XXZ model system Hamiltonian with a transverse field,
\begin{align} \label{kagomeham}
    \ham =& J\sum_{\braket{i,j}} \left( X_i X_j + Y_i Y_j \right) \nonumber \\
    &+ J_Z \sum_{\braket{i,j}} \left( Z_i Z_j \right)
    + B_Z \sum_{i=1}^{3\times N_X \times N_Y} Z_i,
\end{align}
where the sum is over all the nearest neighbors, $\braket{i,j}$ pairs (see \cref{fig:group_kagome}).
The total number of sites for $N_X$ unit cells in the $X$ direction and $N_Y$ in the $Y$ direction is $\{3 \times N_X \times N_Y\}$.
The interaction strength $J$ sets our energy scale, and $J$ is positive, which implies the antiferromagnetic nature of the system. 
We set $B_Z = 0.2$, and will study the energy spectrum as a function of $J_Z$. 

The triangular lattice structure can lead to geometric frustration in the antiferromagnetic regime, leading
to degeneracy and level crossings, both of which can be seen in \cref{fig:group_kagome}.
When $B_Z=0$ there is phase transition near $J_Z/J = -0.5$ 
where the ground state is highly degenerate and is double degenerate for $J_Z/J > -0.5$ \cite{changlani_macroscopically_2018}.
In the presence of a finite external field $B_Z$, 
the degeneracy at the phase transition point spreads out, 
where energy levels are close but differ by some finite energy values (see \cref{app:kagome model}).
These energy gaps between the ground state and the first excited state change with the values of the parameter $J$, as shown in the \cref{fig:kagome_energyDiff}. 
For smaller values of $J$, we observe higher energy gaps before the phase transition and vanishing values after. 
This trend becomes the opposite as we increase the interaction strength $J$. 
As discussed before, non-vanishing energy gaps are crucial for the convergence using methods like ITE. 
Getting a basis for these $J$ values using ITE is challenging due to vanishing gaps for some $J_Z$ values, before and after the phase transition point.
We will address this issue by making use of the middle point at $J=1$ where energy gaps are mostly non-zero throughout the $J_Z$ range (red squares in \cref{fig:kagome_energyDiff}),
and make a basis by truncated ITE at Hamiltonians with $J=1$ and $J_Z$ marked with green diamonds in \cref{fig:group_kagome} part (a) inset. 
We will use this basis for a total of $225$ Hamiltonians with other parameter values at $J>1$  and $J<1$.

In \cref{fig:group_kagome} part (a), we show the spectrum for $J=0.5$ and $B_Z=0.2$ using a basis made from $J=1.0$ and $B_Z=0.2$. 
Basis vectors are obtained from implementing ITE at ($k_p = 6$) equally sampled parameter points (green diamonds) in $J_Z$ with $d\tau=0.2$ and $\taumax=2 $ starting with a Haar random state.
Values with the green diamonds in the color plot are errors in the basis states used in the subspace. 
Now we will use the same basis for more parameter values changing $J \in \{0.1,0.2,..1.4,1.5\}$. 
To compare with the imaginary time evolution for each target point, 
panel (b) plots the fidelity of the ground states corresponding to the target points in panel (a) for $J= 0.5$,
and the fidelity if we implement just $10$ ITE steps at those target point Hamiltonians without using the subspace formalism.
Panel (c) in fig.\cref{fig:group_kagome} plots the relative error in energies solved using the same basis for different parameter points. 
Panel (d) is plotted using only ITE at each parameter point to get the ground state energy with $d\tau=0.2$ and $\taumax=2.0$. 
That is $10$ imaginary time steps for each parameter point.
The horizontal dashed lines at $J=0.5$ in panels (c) and (d) correspond to the states shown in panels (a) and (b). 
In comparison, in (c) we use $(10\times6=60)$ iterations of ITE for getting energy at $(15\times15=225)$ parameter points, and in (d) we use $(15\times15\times10=2250)$ iterations of ITE. 
The resource requirements for the implementation of non-unitary imaginary time evolution operator $(e^{-\tau \ham})$ 
scales with the total time of evolution, or the number of imaginary time steps \cite{motta_determining_2020, lin2021real}.
Using the EC framework, we can significantly reduce the number of time steps required, hence saving computational resources.

Hamiltonians with vanishing energy gaps and degeneracies are in particular more difficult to get to the ground state.
The phase transition point $J_Z/J = -0.5$ (brown dashed line in the color plot in \cref{fig:group_kagome}) is one such difficult point, whereas, at $J=J_Z$ it is easy to converge using ITE.   
The squares with the highest error in the middle plot (c) are at phase transition points where $J_Z/J = -0.5$. 
Where the ground state is highly degenerate which can be seen in the energy spectrum plot.

\section{Noise resilience}

In practice, since the currently available quantum hardware is noisy, we require algorithms to be noise resilient in order to yield useful results on a quantum device \cite{preskill2018quantum, Huang2023near, oftelie2021simulating}.
We already have some ``noise'' in the subspace in the form of excited states,
arising from truncated state preparation methods;
however, quantum devices introduce hardware noise as well, 
requiring us to analyze how additional noise will affect the algorithm. 
In this section, we will demonstrate that the improvements obtained using EC survive the addition
of hardware noise. We implement a noise model and simulate the effect of introducing noise in the ASP for a 5-site XY model Hamiltonian discussed previously in Eq.~\eqref{XYham}.

Implementation of an algorithm on quantum hardware requires mapping of the time evolution operators on a quantum circuit. The presence of noise turns the otherwise unitary evolution gates into quantum channels;
however, defining the noise channel precisely is difficult as it depends on several factors. 
For this reason, a variety of noise models have been suggested for the simulation of different hardware environments \cite{harper_efficient_2020, Georgopoulos_2021}.
Our goal is to investigate the effect of noise on EC, here we limit our consideration
to a general noise model with a single noise variable \cite{zanetti_simulating_2023, Barnum1998information}.

In this noise model, each gate $R$ is followed by a noise gate \cite{isakov2021simulations}, whose amplitude is taken from 
the absolute value of a normal distribution
with a standard deviation $\sigma$ that can be varied to control the noise.
These noise gates will implement the noise error channels for both single-qubit and two-qubit gates.
We will explore the effective action of gates followed by noise gates parameterized by a single noise parameter $\sigma$.
Let $R_i$ denote the gate $R$ at qubit $i$ and $I_i$ is the identity operator on $i$ qubits.
For simple Pauli gates $R\in \{X,Y,Z\}$, we have $R_iR_i = I_i$. 
The effect of a noise gate for a single qubit gate $R_i$ with probability $p$ followed by gate $R_i$ is equivalent to applying $\sqrt{(1-p)} R_i+\sqrt{p}I_i$. 
This accounts for gate imperfection and noise channels; it represents a bit-flip noise channel when 
$(R=X)$, and a phase-flip noise channel when $(R=Z)$. 
Similarly, a two-qubit gate $R_1R_2$ will have noise gates $R_1$ with probability $p1$ and $R_2$ with $p2$. 
The effective operation will thus become (after normalization), $(\sqrt{(1-p1-p2-p1p2)} \; R_1R_2 + \sqrt{p1} \; R_1 + \sqrt{p2} \; R_2 +  \sqrt{p1p2}\; I_2)$.
For the 5-qubit system considered here, for every single and two-qubit gate, we will have effective gates as:
\begin{align}\label{eq:Noise_model}
    R_1 \rightarrow& \sqrt{(1-p)} \; R_1 + \sqrt{p}  \; I_5 \nonumber \\ 
    R_1R_2 \rightarrow& \sqrt{(1-p1-p2-p1p2)} \; R_1R_2 + \nonumber\\  
    		 &\sqrt{p1} \;R_1 + \sqrt{p2} \; R_2 + \sqrt{p1p2} \; I_5
\end{align}

\begin{figure}[t]
    \centering
    \includegraphics[clip= true, trim=0cm 0cm 9cm 11.2cm, width=0.48\textwidth]{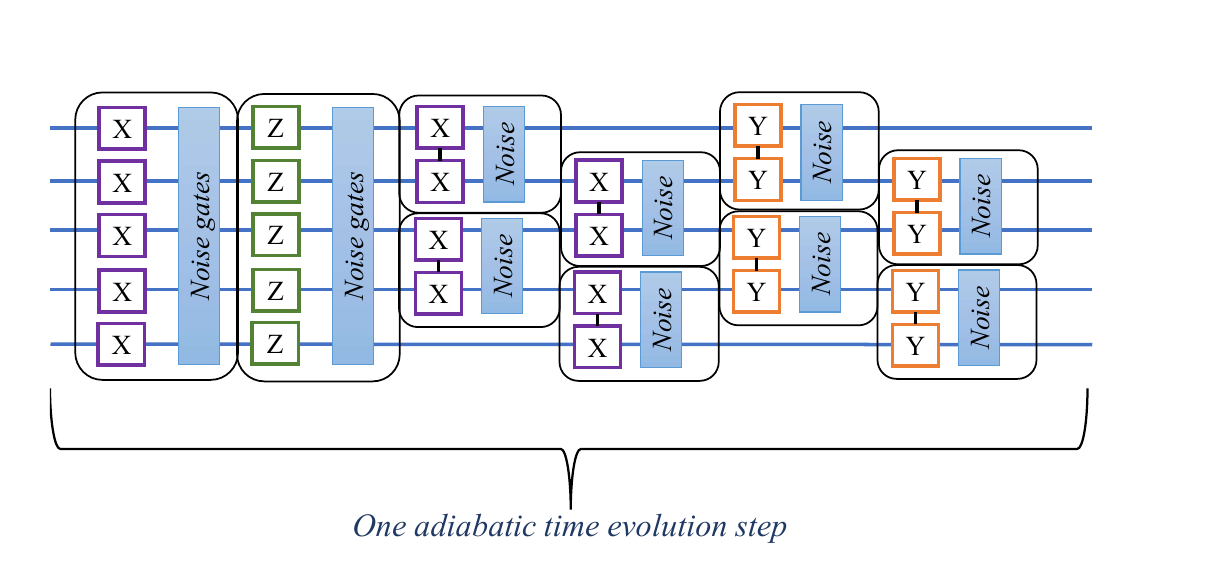}
    \caption{One step of adiabatic time evolution for a 5 site XY model Hamiltonian with noise}
    \label{fig:ATE_circuit}
\end{figure}

We implement this noise model in the ASP procedure to generate the subspace basis states.
Each time step of the adiabatic evolution for this system involves a Trotter decomposition 
as per \cref{XYham}. 
For all the terms in the Hamiltonian, each is followed by a noise term (see \cref{fig:ATE_circuit})
that changes the Hamiltonian at that time step to $(\Bar{H} = \ham + \ham_{noise})$,
i.e. we replace the Hamiltonian in the time evolution operator $e^{-i\mathcal{H} dt}$ by $\Bar{H}$.
The probabilities in each noise gate are sampled from a distribution, and we perform an average over
an ensemble of 500 trajectories.
For each trajectory, we calculate the projected Hamiltonian ($\hammat$) and overlap matrix ($\ess$) elements;
we then bootstrap the sampled data to get the average matrix element values.
(The bootstrapper error bars for the matrix values are smaller than $\epsilon < 10^{-3}$).
The Hamiltonian and overlap matrices, which now contain the averaged elements, are then solved by the
GEP to obtain ground state energy values.

\begin{figure}[htpb]
    \centering
    \includegraphics[width=0.48\textwidth]{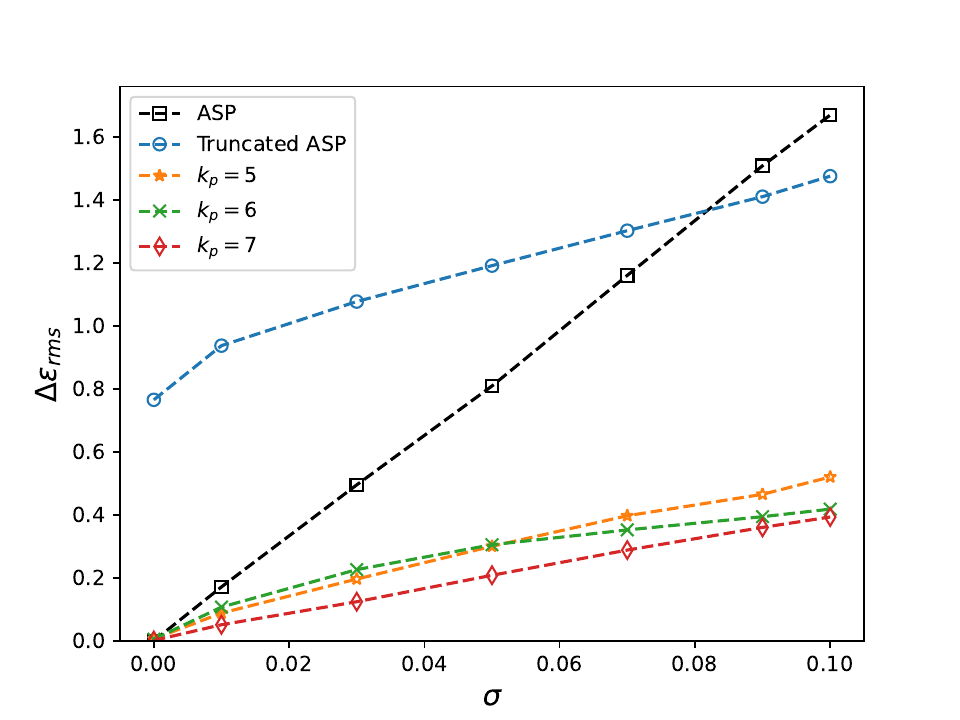}
    \caption{Noise analysis for 5-site XY 1D model with just ASP and with EC for the different number of basis vectors $k_p$ while varying noise $\sigma$ on the x-axis. The y-axis denotes the RMS error ($\Delta\varepsilon_{rms}$) for all the target points. }
    \label{fig:noise_plot}
\end{figure}

In the \cref{fig:noise_plot}, we plot the error in the ground state energy $\Delta\varepsilon_{rms}$ (\cref{eq:RMS}) as a function of the noise parameter $\sigma$ for $20$ target points of the XY model.
We compare the use of ASP ($\tmax=75$) and truncated ASP ($\tmax = 7.5$), both with $dt=0.05$,
and the effect of correcting the truncated ASP using EC with $k_p \in \{5,6,7 \}$ basis vectors.
In all cases, increasing the gate noise corresponds to an increase in  $\Delta\varepsilon_{rms}$.
The ASP circuit has the largest depth ($1500$ time steps) and thus the error increases most rapidly.
The truncated ASP results have a more modest depth ($150$ time steps) and thus show a more gentle
increase, but at the cost of a large error at $\sigma=0$ --- this is of course expected because the $\tmax$
is insufficient for an adiabatic evolution. Using EC on top of the truncated ASP fixes the issue at $\sigma=0$,
but also improves the situation for finite $\sigma$.  
For comparison, we repeat the calculations with an increased number of basis vectors in the subspace from $5$ to $6$ and $7$.   As expected,
increasing the number of basis vectors does help to reduce the error ($\Delta\varepsilon_{rms}$) somewhat.
As discussed above, we select equally spaced values in $B_Z$, which for 
$k_p = 6$ results in some of the training points being very close to the level crossings similar to the plot \cref{fig:group_XY5} (f), which causes the error to increase.
These results indicate that even in the presence of noise, making use of this subspace method improves the results.

While increasing the number of training vectors (or subspace basis states) can help to
reduce the error, this eventually reaches a barrier.
Adding more basis vectors
increases the chances of over-spanning the subspace, making the condition number $(C_n)$ of the overlap matrix very large, and yielding it non-invertible (see \cref{eq:GEP}). 
This is one of the major problems with subspace methods (\cite{cortes2022quantum, zhang2023measurementefficient, kirby2024analysis}).
To address this issue, there are several classical post-processing techniques used to manipulate the overlap matrix, such as thresholding \cite{Epperly2022theory}.
This procedure involves finding the singular value decomposition of the overlap matrix, followed by eliminating the singular values that are smaller than a chosen threshold ($\alpha$). 
The vectors corresponding to the singular values less than $\alpha$ are discarded from the overlap matrix and the projected Hamiltonian matrix.
This technique can help in reducing the condition number but sometimes we can lose some relevant information in the process. 
Let us analyze the effect of noise on condition number and thresholding for our test case. 

\begin{figure}[ht]
    \centering
    \includegraphics[width=0.48\textwidth]{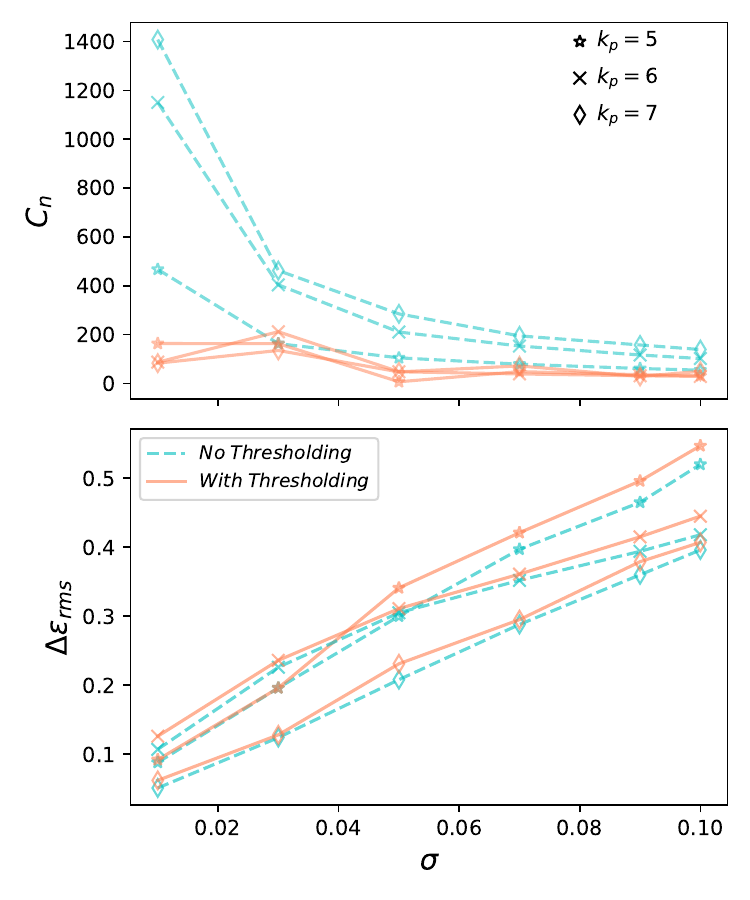}
    \caption{ Condition number ($C_n$) and RMS error ($\Delta\varepsilon_{rms}$) analysis with and without Thresholding for changing noise ($\sigma$)}
    \label{fig:thres_noise}
\end{figure}

We plot the condition number ($C_n$) and error ($\Delta\varepsilon_{rms}$) before and after thresholding with different noise for the above-mentioned data in \cref{fig:thres_noise}. 
Thresholds ($\alpha$) are chosen empirically as the noise ($\sigma$) is varied,
for $\sigma < 0.04, \alpha =0.015$; $\sigma < 0.08, \alpha = 0.05$ and $\sigma > 0.08, \alpha = 0.09$. 
From the plot, we observe that the condition number increases as $k_p$ is increased from $5$ to $7$ and the RMS error decreases.    
However, as $\sigma$ increases, the condition number decreases;
noise helps keep the overlap matrix invertible. 
Using the thresholding technique does help in reducing the condition number as shown in the first plot in \cref{fig:thres_noise}, but for larger $\sigma$, it is unnecessary
as the condition number is already under control, and using it anyway somewhat increases the RMS error as seen in the second plot.

To understand why noise in EC basis helps keep the condition number in check while introducing more errors, 
we need to look at the change in the EC basis with noise. 
The EC basis is generated using the low-energy states, where adding noise introduces other higher-energy states. 
These randomly added states decrease the overlap of the basis with other states, reducing the condition number.
Contrariwise,
in other truncated methods such as Krylov, the subspace basis is already
somewhat spread out, having contributions from all the states. 
Here, adding noise can increase the contributions of some of the states thus increasing the condition number.
(See Appendix \ref{Noise analysis} for an example.)

\section{Discussion}

In this work, we have used previously proposed eigenvector continuation as a subspace method \cite{francis2022subspace},
where continuity of the ground state of the system Hamiltonians varied by some order parameter is explored.
In contrast to previous work, where exact eigenstates
were needed to generate the subspace,
we apply it to
the subspace generated by 
truncated time evolution techniques (ASP and ITE) and the variational quantum eigensolver (VQE) to generate a subspace.
These truncated algorithms incorporate a number
of low-energy states, which can be disentangled using
EC, thus gaining better convergence while using fewer resources.

We used various spin Hamiltonians with different ground state energy spectrum features to showcase the method,
including 
the XY model with and without transverse field whose ground state manifold spans orthogonal symmetry sectors, 
the XXZ 1D model with a transverse field which has small energy gaps between ground state and the first excited state, and the
XXZ Kagome 2D model which has phase transitions and ground state degeneracies. 
However, this method is not limited to spin systems;
the eigenvector continuation formalism has been previously used for other systems in nuclear physics, 
and quantum chemistry among others \cite{yapa_volume_2022, Mejuto_Zaera_2023, duguet2023eigenvector}, and
the truncated state preparation method proposed in this work can also be used for these systems. 

All the findings shown in this work were performed on a classical simulator. 
Even here, 
compared to the classical full-time-evolution methods to get the ground states, 
using EC with truncated time evolution can be much faster.
For quantum simulations, however, it can be challenging to obtain the matrix elements of the projected Hamiltonian $\hammat$ and overlap matrix $\ess$ [Eq.~\eqref{eq:GEP}]. 
One can evaluate these elements by performing multi-qubit controlled operations to an ancilla, e.g., the Hadamard test method \cite{somma2002simulating}. 
These operations are known to be troublesome for current QPUs, as they introduce significant error. 
Other approaches for measuring overlap terms have been recently proposed to avoid the need to use the traditional Hadamard test,
including a cheaper implementation when using ASP for producing the basis states \cite{shen2023real}, a SWAP test that could be cheaper than the Hadamard test \cite{huggins2020non},
and an approach that relies on a reference state\cite{cortes2022quantum}.
Moreover, algorithms for calculating the overlap of two quantum states efficiently on a quantum device remains an open research area.

While the calculation of the projected Hamiltonian and subspace matrices can involve otherwise expensive calculations, the number of these is limited.
Because we use the same basis for all the target Hamiltonians, the overlap matrix elements need to be calculated once.
Similarly, for the projected Hamiltonian matrix, usually, we do not need to repeat the measurements for each target Hamiltonian, since all the Pauli operators of the Hamiltonian are measured only once per basis set and only the coefficients are changed to calculate the target Hamiltonian \cite{francis2022subspace}.
Due to the structure of the matrices, the total number of circuits required scales quadratically in the number of basis states $k_p$: $k_p(k_p-1)$ for the overlap matrix and $k_p(k_p+1)$ for each of the Pauli operators in the projected Hamiltonian.

Overall, we believe that this method will be of high
value for quantum computing in the near term, and into
the early fault tolerant era.  In both cases, circuit
depth will come at a premium, and any algorithm that
can reduce the depth and provide some noise resilience while maintaining or reducing the
error can have an impact.

\section*{Acknowledgements}
This work was supported by the
Defense Advanced Research Projects Agency (DARPA)
under Contract No. HR001122C0063. 
Any opinions, findings, conclusions or recommendations expressed in this material are those of the author(s) and do not necessarily reflect the views of the Defense Advanced Research Projects Agency. The variational quantum algorithm results were supported by the U.S. Department of Energy, Office of Science, Office 
of Advanced Scientific Computing Research under Award Number DE-SC0025623.

\bibliographystyle{apsrev4-2}
\bibliography{kemperlab}

\clearpage

\renewcommand\thefigure{A\arabic{figure}}  
\renewcommand{\theequation}{A.\arabic{equation}}
\setcounter{figure}{0}

\appendix

\section{Eigenvector continuation}\label{app:EC}
For a model Hamiltonian that depends on certain parameters, what is the change in the ground state and eigenenergy resulting from a small change in the parameter; Analysis of this question says the change should be continuous and tractable. 
Eigenvector continuation is based on the same intuition. Across a parameter range, the ground states for each Hamiltonian varied by that parameter are spanned by a few low-energy vectors. 
We will use this idea to form a subspace containing the ground states across the parameter range. This will let us use the subspace diagonalization for each such Hamiltonian using the subspace thus formed. The execution of this method is shown below.
The aim is to find the ground states for a set of Hamiltonians $\mathcal{H}_{target}$ varied by a parameter $a$. 
\begin{equation*}
    \mathcal{H}_{target} = \{  H(a_1), H(a_2) \dots H(a_n)\}
\end{equation*}

From the set, choose $k_p < n$ points in parameter space and corresponding Hamiltonians, we call this a training set $\mathcal{T}$. Ground states at these points will be used as basis vectors in the subspace.
The number and position of these Hamiltonians can be chosen based on prior knowledge about the system, the presence of symmetries and the size of the Hamiltonians.  
\begin{equation*}
    \mathcal{T} = \{  H(k_1), H(k_2) \dots H(k_p)\}
\end{equation*}
We will use the quantum state preparation techniques including time evolution methods such as adiabatic state prep, and imaginary time evolution for getting the ground states. 
Here since the requirement is for low-energy subspace, running a quasi-ground state preparation is enough. The states obtained from these methods would have a finite overlap with other excited states. These vectors form a basis $\mathcal{B}$ for subspace diagonalization. 
\begin{equation}\label{basis_ec}
    \mathcal{B} = \{  \phi_1, \phi_2 \dots \phi_{k_p}\}
\end{equation}

\section{QITE} \label{app:QITE}
\begin{figure}
    \centering
    \includegraphics[width=0.48\textwidth]{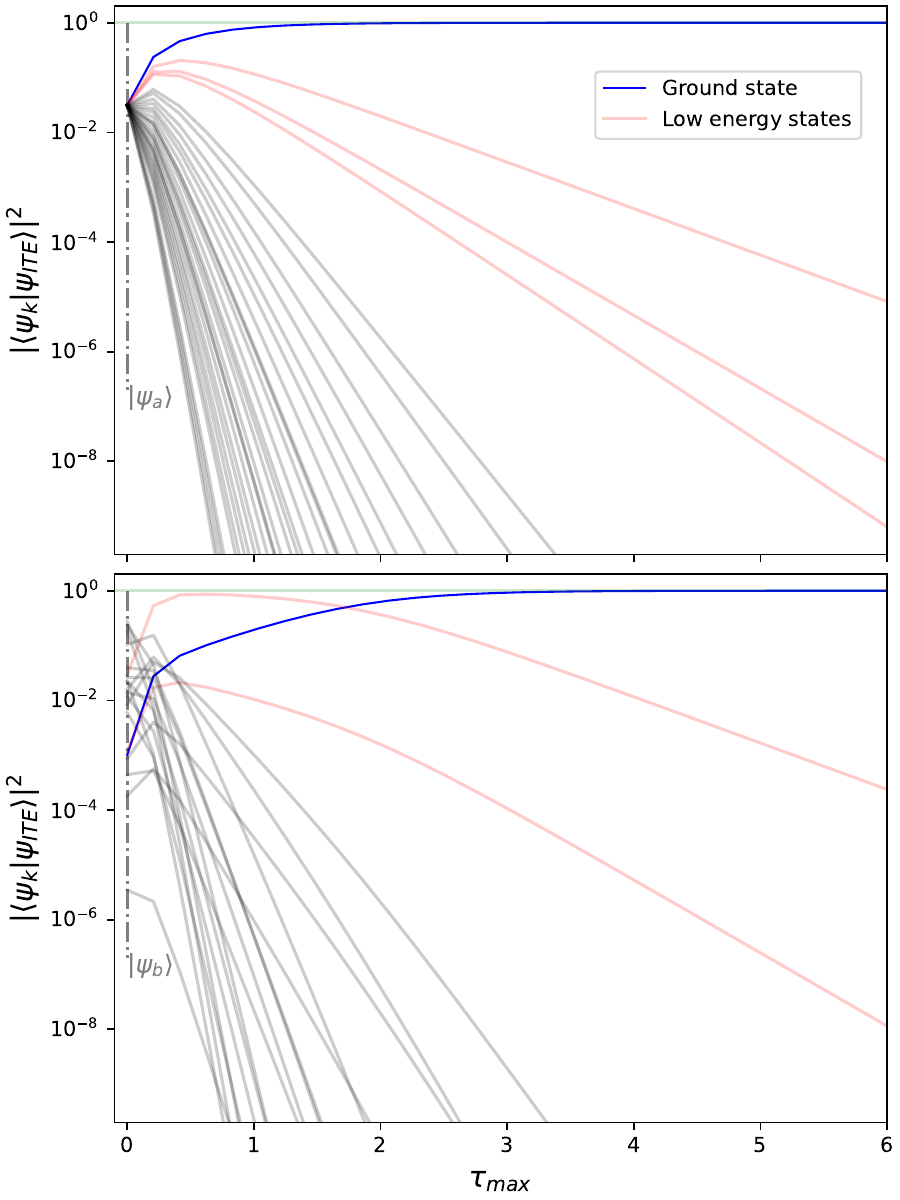}
    \caption{Change in the imaginary time-evolving state overlaps with the eigenstates of the XY 1D 5-site system for every step of evolution. The initial state $\ket{\psi_a}$ is an equal superposition of all the eigenstates of the system and $\ket{\psi_b}$ is an equal superposition of all the 5-qubit computational basis states.  }
    \label{fig:ITE evolution}
\end{figure}

Imaginary time evolution involves an evolution of the initial state with imaginary time, which is $e^{-H_t\tau}$.
ITE requires the initial state to have some finite overlap with the ground state. In practice starting with a Haar random state, the probability of that state being completely orthogonal to the ground state is very low. 
Let the starting state be $\ket{\psi_0}$. The state can be written as a superposition of all the eigenstates($\ket{e_i}$).
The application of evolution term $e^{-H_t\tau}$ where $H_t$ is the target Hamiltonian, on the initial state followed by normalization, increases the overlap with eigenstate corresponding to the low(est) eigenvalue (highest eigenvalue for $e^{H_t\tau}$) and overlap with eigenstate corresponding to the high(est) eigenvalue decreases. Iterating this many times (large $n$) gives maximum overlap with the ground state or lowest eigenvalue state (say, $\ket{e_0}$) and all the other coefficients get close to zero. 
\begin{align*}
    \ket{\psi_0} &= \sum_{i=0}^{N-1} c_i \ket{e_i} \\
    \ket{\psi_i} &= \frac{e^{-H_t d\tau}\ket{\psi_{i-1}}}{\bra{\psi_{i-1}}e^{-2H_t d\tau}\ket{\psi_{i-1}}} \\
    \ket{\psi_n} &\sim \ket{e_0}; \; n>>1
\end{align*}
The \cref{fig:ITE evolution} plots how a state evolves under imaginary time.  
Starting with an initial state that is an equal superposition of all the eigenstates $\ket{\psi_a} = \sum_{k=1}^{2^N} 1/\sqrt{2^N} \ket{\psi_k} $, overlap with low(est) energy (red) eigenstates increases then slowly starts to decrease. 
In contrast, overlap with higher energy eigenstates (grey) vanishes rapidly. 
Getting rid of those low-energy states takes the longest total time ($\taumax$).
However, preparing this state is not practical.
Instead, we use the state $\ket{\psi_b} =  1/\sqrt{2^N} \ket{1,1,\dots 1}$, which is an equal superposition of the computational basis states in the second plot.
Similar behavior of low-energy states can be observed where the ground state takes longer to converge. 

 Since the evolution term ($e^{-H\tau}$) is non-unitary it makes implementation of ITE on quantum devices non-ideal. 
 There exist various methods for this purpose \cite{lin2021real, motta_determining_2020}, however, they are very costly. Therefore limiting the number of evolution steps is preferable. Using EC number of steps needed can be reduced.

\section{ASP}\label{app:ASP}
Adiabatic state preparation (ASP) is used when the ground state at some parameter value is known or is easy to find and the ground state at some other parameter value is needed. To find the ground state at a target value, we start with the known ground state at some parameter value say $\ket{\psi_i} = \ket{\psi_{gs}(p_i)}$ and slowly (adiabatically) evolve the state changing the Hamiltonian to the target parameter value $(p_f)$. 
If the evolution is N step process, and the change in parameters is 

\begin{equation*}
    dp= (p_f-p_i)/N
\end{equation*} 
then each adiabatic time step will be, 
\begin{equation*}
    \ket{\psi_j} = e^{-iH(p_i +j *dp) dt} \ket{\psi_{j-1}}
\end{equation*}
and the total evolution will be, 
\begin{equation}
    \ket{\psi_f} =  e^{-iH(p_f) dt} e^{-iH(p_f-dp) dt} \dots e^{-iH(p_i+dp) dt} \ket{\psi_i} 
\end{equation}

Here if N is not large enough, the evolution can accumulate diabatic excitation. 
Furthermore, if the value of $dt$ is not small enough, it can increase the Trotter error. We thus require small $dt$ and large $N$ for the evolution to be exact, which however requires large computational resources. 
The total time of evolution required to follow the lowest eigenstate depends on the energy gap between the ground and the first excited state. If energy gaps are small, there is a higher probability of state mixing while evolution and even smaller time steps would be required. For situations where there are points of no energy gap, degenerate points, or level crossings, ASP fails to detect the change in the ground state. And the resultant state is not the ground state.

In such systems, a symmetry-breaking term is introduced in the Hamiltonian which breaks the level crossings in the energy spectrum and introduces an energy gap. For small field strengths, a very long evolution is required. For a toy model, TFXY model with X-field to break symmetry (\ref{XYham}), we will analyze this situation in detail.

\begin{figure}[htpb]
    \centering
    \includegraphics[width=0.48\textwidth]{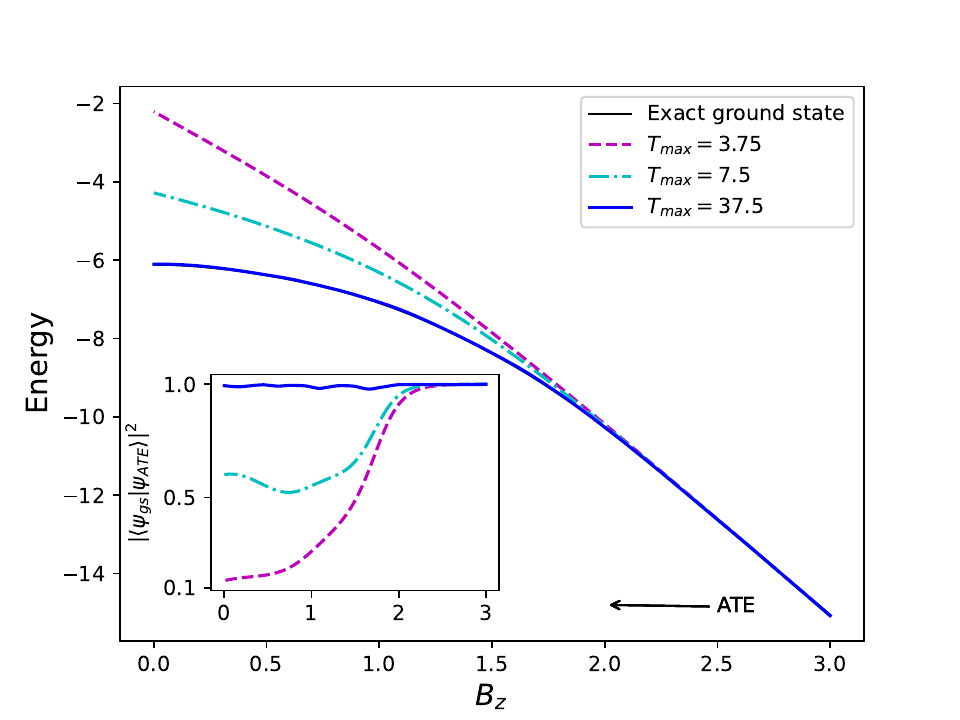}
    \caption{Adiabatic time evolution for a 5-site XY model with parameters $J=1, B_X = 0.2$, $dt=0.05$ with different $T_{max} = [3.75, 7.5, 37.5]$ and changing $B_Z$ and (inset) corresponding ground state fidelity}
    \label{fig:ATE_evolution}
\end{figure}

Consider implementation of adiabatic state preparation for a 5-site model, starting with $B_Z= 3.0$ in \cref{fig:ATE_evolution}. For the same $dt=0.05$ and changing the number of adiabatic time steps by changing $\tmax$, the initial ground state at $B_Z=3.0$ evolves to different states at $B_Z=0.0$. The fidelity of the ground state with the evolving state remains $1$ only for large enough $\tmax = 37.5$. This would require $(n=750)$ number of adiabatic time evolution steps. If we evolve for lesser $\tmax$, say $3.75$ this would take $10\%$ of the evolution steps but will not capture the changes in the ground state, as shown in the figure. This quasi-adiabatic evolution is used in the text with EC.    

\section{Models}

\subsection{XY model}\label{app:XY model}
In the paper, a transverse field XY model with a uniform longitudinal field is used as a toy model. The model involves in-plane interactions in X- and Y- directions, here nearest neighbour interactions are considered and a universal field in the Z-direction of strength $B_Z$. In-plane interaction strength $J$ is kept constant and $B_Z$ is varied to plot an energy spectrum.
Hamiltonian (restated) is given by: 
\begin{align} 
    \ham =& J\sum_{i=1}^{N-1} \left( X_i X_{i+1} + Y_i Y_{i+1} \right) \nonumber \\
    &+ B_Z \sum_{i=1}^N Z_i
    + B_X \sum_{i=1}^N X_i,
\end{align}

\begin{figure}[htpb]
    \centering
    \includegraphics[width=0.48\textwidth]{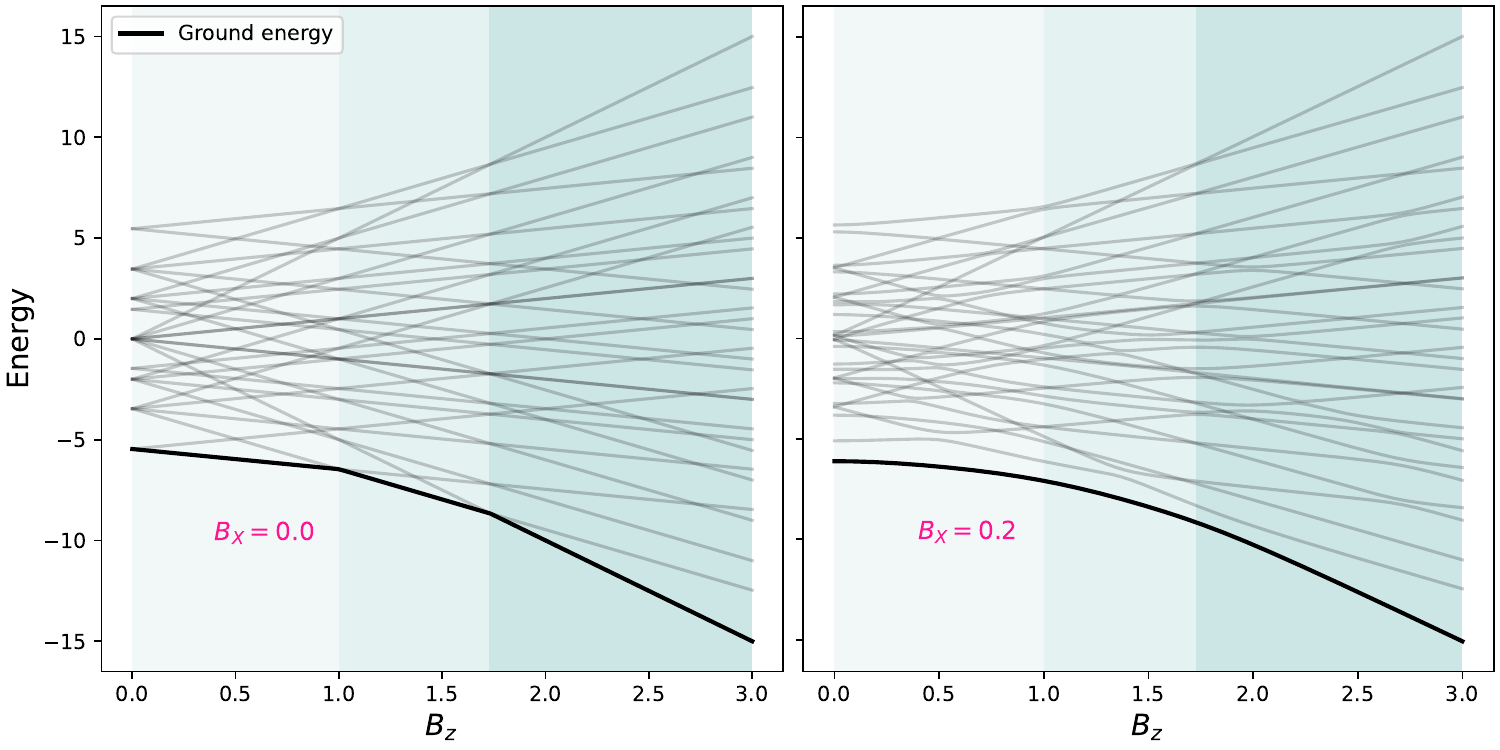}
    \caption{XY 1D chain system energy spectrum for $J=1$, varying $B_Z$ and staggered $B_X \in \{0.0,0.2\}$. Level crossing split can be observed in the plot on the right.}
    \label{fig:XYspectrum}
\end{figure}

\begin{figure}
    \centering
    \includegraphics[width=0.48\textwidth]{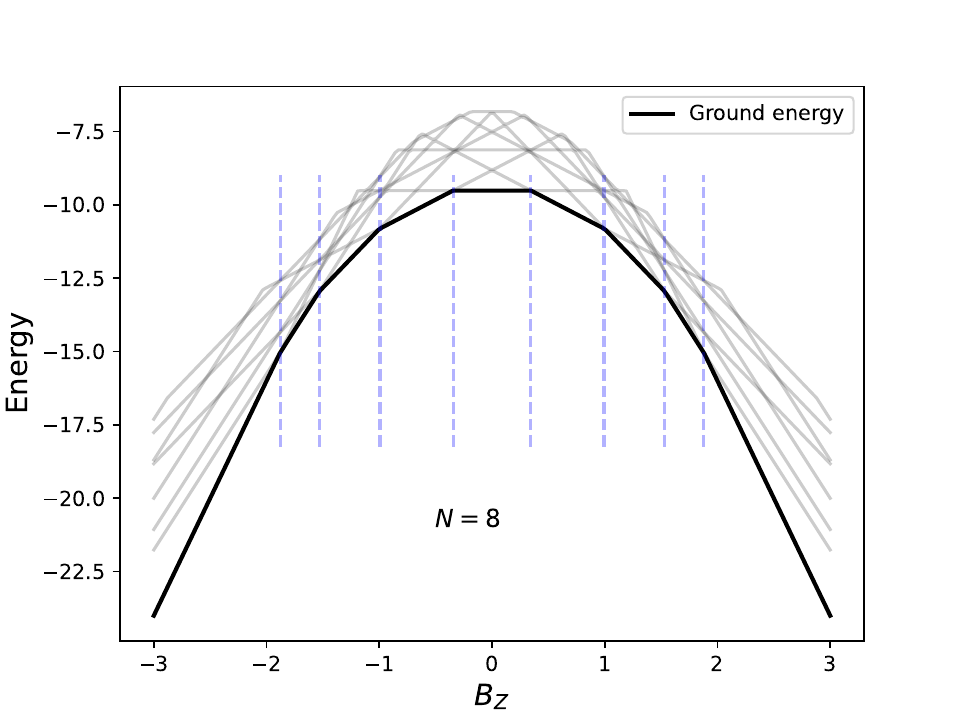}
    \caption{Scaling of number of symmetry sectors in 1D transverse field XY chain is $O(N)$. Here, for instance, we plot the energy spectrum for $N=8$ with varying transverse field $B_Z$. Blue dashed lines show points of level crossing.}
    \label{fig:XY8}
\end{figure}

Without a staggered X- field, the ground state is protected for a range of varying $B_Z$ due to the presence of magnetization symmetry sectors. 
For particular values of $B_Z$, the ground state changes to some excited state resulting in a point of degeneracy where both states cross called level crossing. 
The ground state before and after these points is completely orthogonal.
For a $N$ site 1D chain system, for each spin flip, we will have a magnetization phase transition making the number of symmetry sectors in 1D transverse field XY model with varying transverse field from $B_Z \in (-\infty, +\infty)$ as $(N+1)$. 
For example we plot $N=8$ energy spectrum with $J=1$, staggered $B_X=0.0$ and $B_Z \in (-3,3)$ in \cref{fig:XY8}.
For the plots in this paper, we consider $B_Z \in [0,B_{Z(max)})$. 
Still, the scaling of the number of symmetry sectors would be linear in $N$.
Adding a uniform field in the X-direction of strength $B_X$ has two significant consequences. 
One is the breaking of level crossings, and the other is the mixing of the lowest eigenstates.
Breaking of level crossing introduces an energy gap between the lowest energy states, which makes the application of ASP possible. 
After adding a staggered $B_X$ field, the ground states are not completely orthogonal unlike one without the field.
This makes the training states mixed as well.

\subsection{XXZ model 1D chain }\label{app:XXZ model}
\begin{figure}[htpb]
    \centering
    \includegraphics[width=0.48\textwidth]{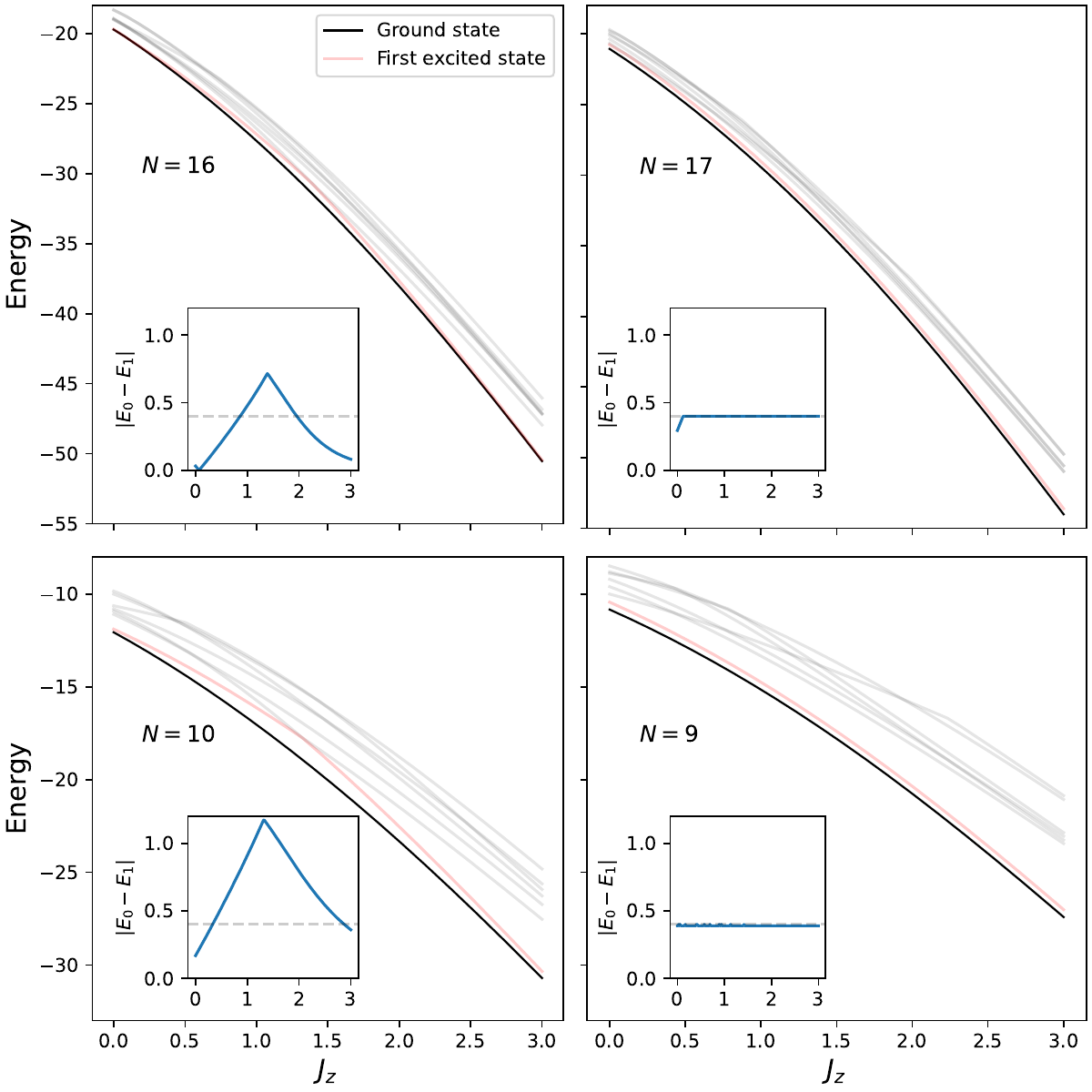}
    \caption{Energy spectrum for XXZ model sites $N \in \{9,10,16,17\}$ with energy gap between the ground state and the first excited state shown in the inset. }
    \label{fig:ener_XXZ}
\end{figure}

Hamiltonian for a XXZ model 1D chain is given by: 
\begin{align} \label{ham:XXZ}
    \ham =& J\sum_{i=1}^{N-1} \left( X_i X_{i+1} + Y_i Y_{i+1} \right) \nonumber \\
    &+ J_Z \sum_{i=1}^N Z_iZ_{i+1} +B_Z \sum_{i=1}^N Z_i
\end{align}
The energy spectrum for this model shows an odd-even pattern which affects the implementation of ASP in \cref{fig:NvsKp}. 
The reason can be seen in the \cref{fig:ener_XXZ} where energy gaps between the ground state and the first excited state remain the same for an odd number of sites ($N=9,17$) at a constant value of ($|E_0-E-1| = 0.4$). 
The slight variation is due to the presence of a finite $B_Z$ field.
For an even number of sites, however, the gap values change. 
For an even number of sites, the gap decreases as the number of sites increases. 
Since the ASP depends on this gap, systems with larger gaps produce more accurate states with ASP. 
For smaller even number of sites($N=10$), most of this gap is larger than odd ones ($N=9$), and for larger $(N=16)$ the gap gets smaller than odd ones for most of the $J_Z$ values. 
ASP takes longer to converge in case of smaller energy gaps and requires a smaller time evolution step. 
Here since the evolution is kept the same, smaller energy gaps show more error.
This explains the odd-even effects observed in the \cref{fig:NvsKp}.

\begin{figure}[htpb]
    \centering
    \includegraphics[width=0.48\textwidth]{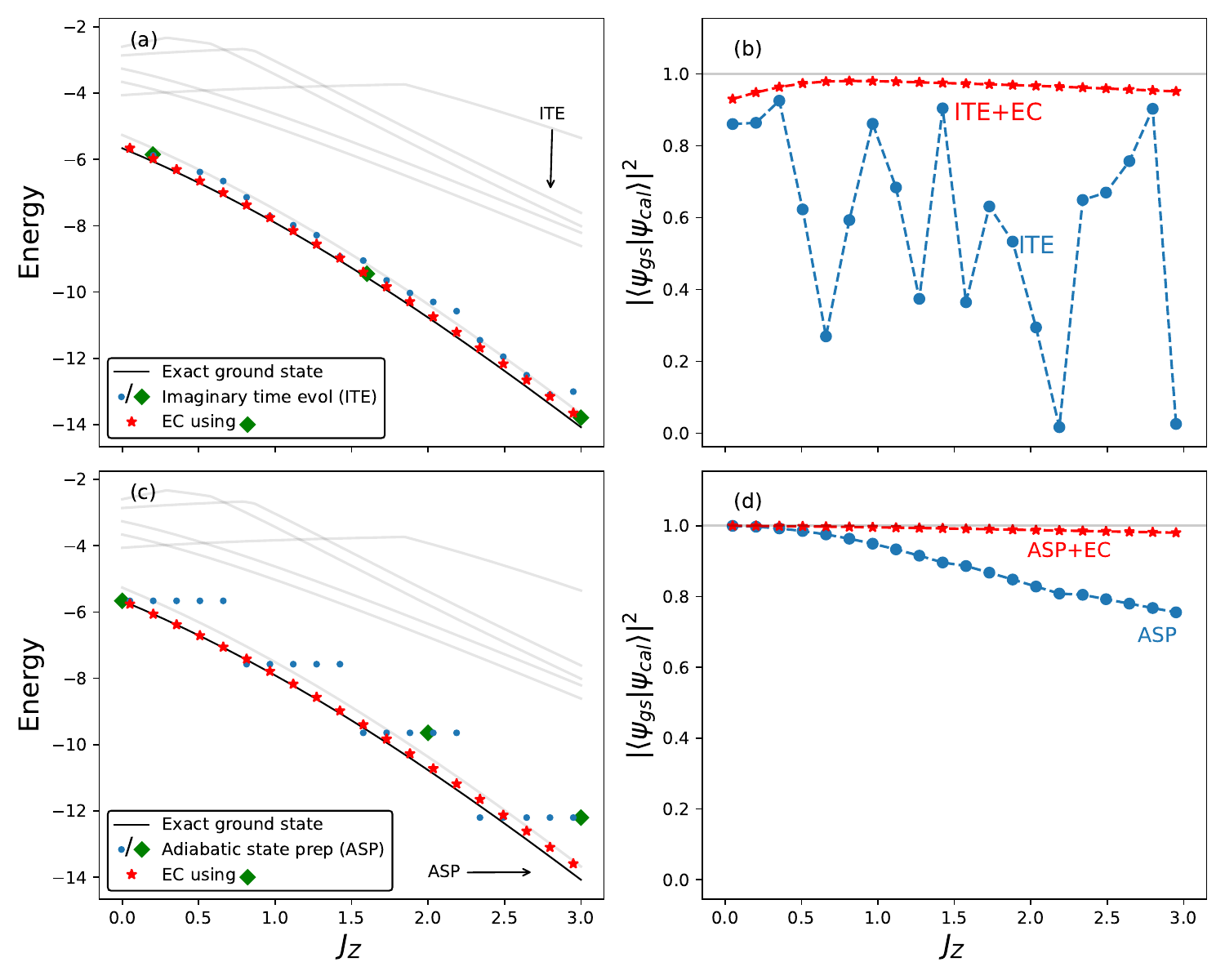}
    \caption{Implementing EC with truncated ASP and ITE for a 1D chain XXZ model 5 site. (a) and (c) plots the ground energy spectrum, (b) and (d) plots corresponding ground state fidelity change.}
    \label{fig:group_XXZ}
\end{figure}

In the \cref{fig:group_XXZ}, truncated ITE and ASP are used to get a basis using and for XXZ 1D chain Hamiltonians \ref{ham:XXZ},
with $J=1$, $B_Z=0.2$, and $J_Z \in (0,3)$ is the varying parameter. 
For truncated ITE in (a) and (b), $d\tau = 0.2$ and $\taumax = 1.2$ starting from a Haar random state. 
And for ASP in (c) and (d), starting from a ground state at $J_Z = 0$, $4$ time steps with $dt=0.1$ and $\tmax = 0.4$ are used to obtain the basis.

\begin{figure}[htpb]
    \centering
    \includegraphics[width=0.48\textwidth]{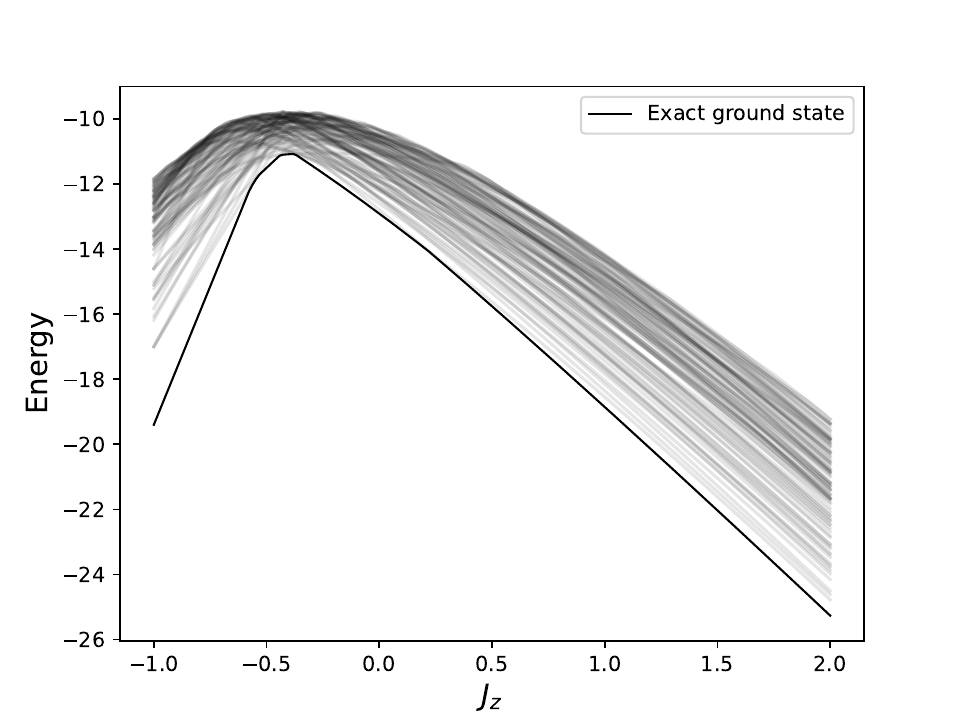}
    \caption{Energy spectrum of a 12-site Kagome Lattice system with $N_X=N_Y=2$, $J=1.0$, $B_Z=0.2$ and varying $J_Z$ showing lowest $100$ energy levels}
    \label{fig:kagome_energyPlot}
\end{figure}

\subsection{Kagome XXZ}\label{app:kagome model}

Hamiltonian for a Kagome XXZ 2D system with lattice structure shown in the \cref{fig:Kagome_lattice} is given by (restated):

\begin{align}
    \ham =& J\sum_{\braket{i,j}} \left( X_i X_j + Y_i Y_j \right) \nonumber \\
    &+ J_Z \sum_{\braket{i,j}} \left( Z_i Z_j \right)
    + B_Z \sum_{i=1}^{3\times N_X \times N_Y} Z_i  \nonumber
\end{align}

where, $J$ is the interaction strength and $J_Z\in (-1,2)$ is the varying parameter. 
The total number of sites would be unit cells ($N_X \times N_Y $) with $3$ sites per unit cell.

\begin{figure}[htpb]
    \centering
    \includegraphics[width=0.48\textwidth]{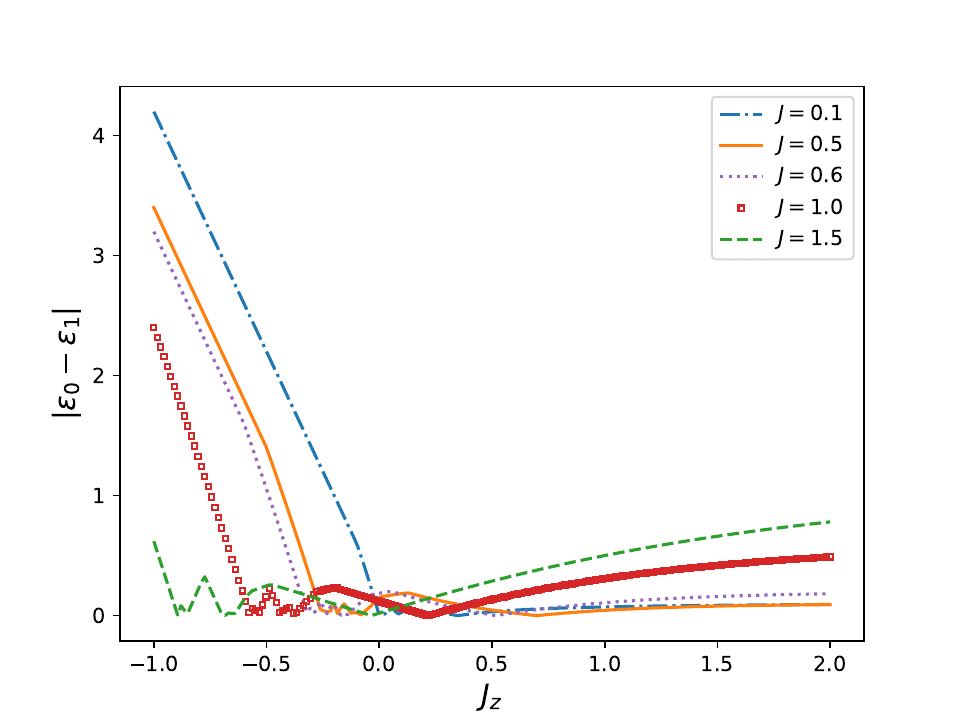}
    \caption{Energy gaps between the ground state and the first excited state for Kagome XXZ Hamiltonians for various parameter values $J \in \{0.1, 0.5, 0.6, 1.0, 1.5\}$ and $J_Z \in (-1.0, 2.0)$.}
    \label{fig:kagome_energyDiff}
\end{figure}

\cref{fig:kagome_energyPlot} plots the energy spectrum for a 12-site model with parameter value $J=1.0$. 
The phase transition point $J_Z/J = -0.5$ is spread out due to the presence of finite field $B_Z= 0.2$. 
This field affects the energy gaps between the ground state and the first excited state which is crucial for the convergence of time evolution methods like ITE. 
In \cref{fig:kagome_energyDiff} we plot the energy differences for different values of the parameter $J$. 
The double degeneracy after the phase transition is broken with increasing energy differences as the value of $J$ increases.
Whereas, before the phase transition region the energy gaps decrease.

\section{Noise analysis}\label{Noise analysis}


\begin{figure*}
    \centering
    \includegraphics[width=0.48\textwidth]{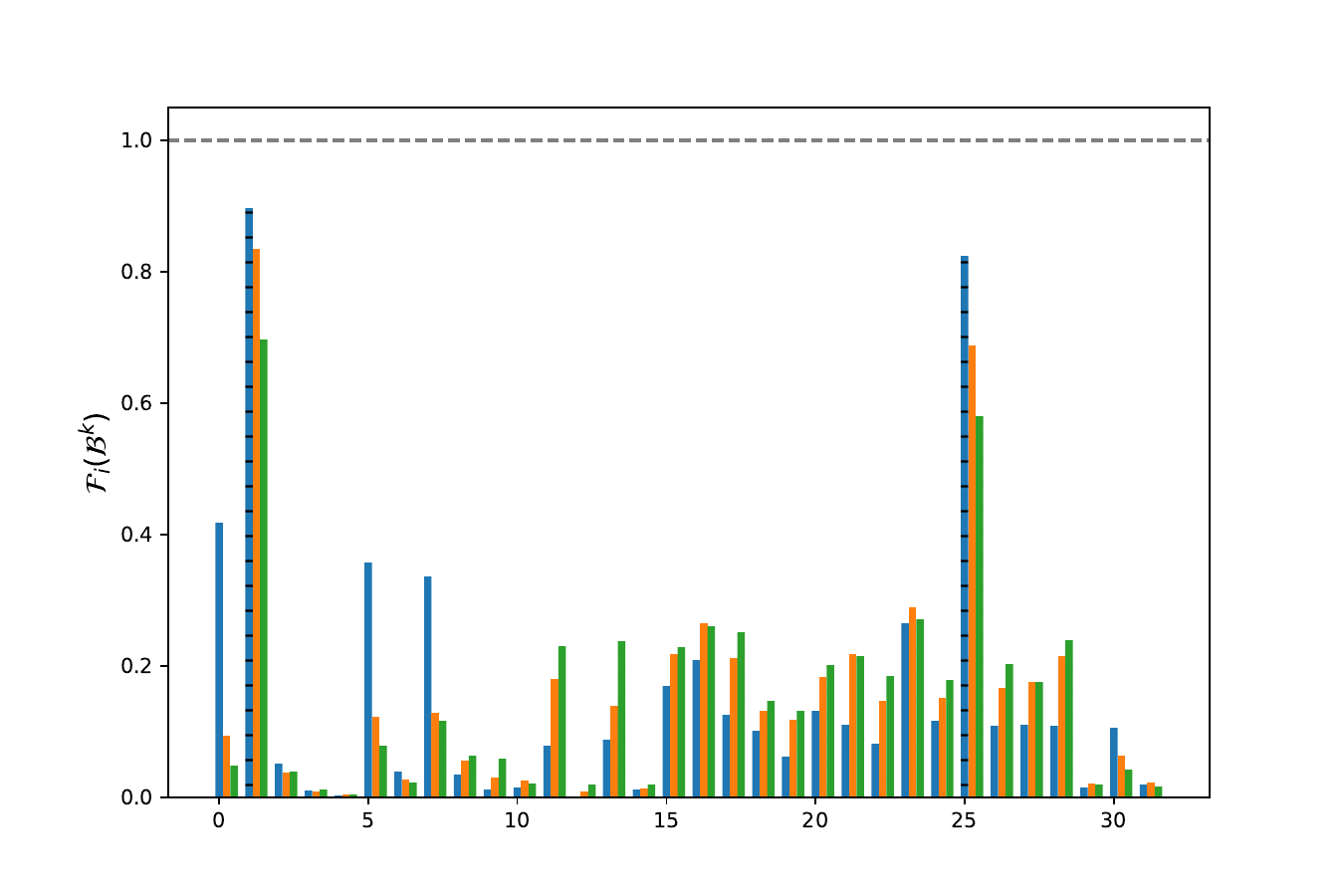}
    \includegraphics[width=0.48\textwidth]{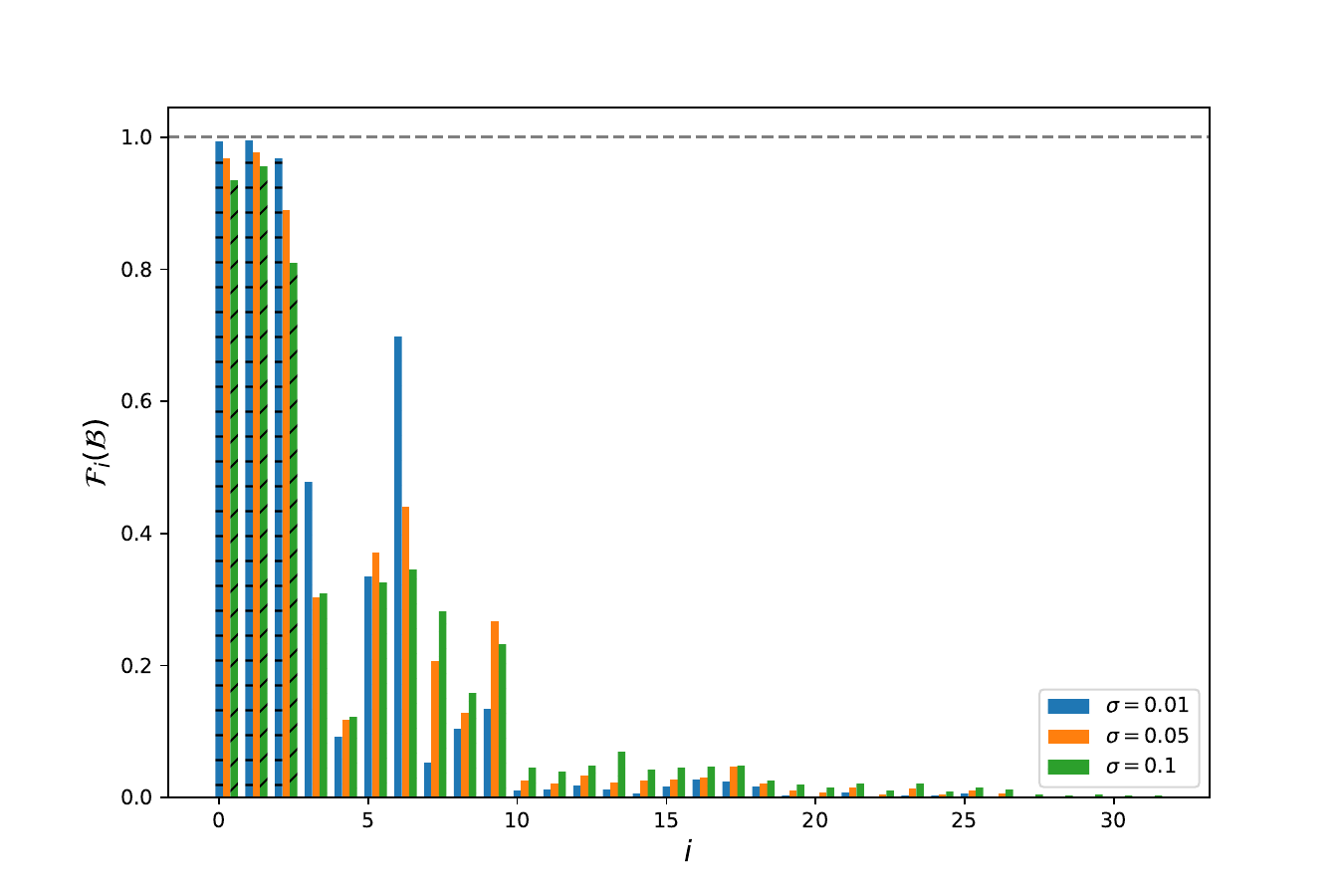}
    \caption{Overlap of the basis with eigenstates with noise $\sigma \in \{ 0.01, 0.05, 0.1\}$ for $k_p = 7 $ basis vectors obtained from Krylov method and truncated ASP method respectively. }
    \label{fig:overlap_basis_noise}
\end{figure*}

We analyze the effect of noise on the subspace basis obtained using truncated ASP and Krylov with $7$ 
basis vectors for target XY Hamiltonian in eq.\ref{XYham} with $J=1; B_X = 0.2$ and, $B_Z = 1.5$.
For getting a basis using Krylov $\mathcal{B}^k$, we use operators $\{H, H^2, \dots\}$ acting on a initial state $\ket{\psi_0}$ prepared using equal superposition of computational basis states.
\begin{equation*}
    \mathcal{B}^k = \{  \phi^k_1, \phi^k_2 \dots \phi^k_{k_p}\}
\end{equation*}

To check the overlap of basis vectors with the eigenstates, we will calculate the fidelity of each basis vector $\phi_i$ with each eigenstate $\ket{e_i}$.
The overlap of the basis with eigenstates are given by: 
\begin{equation*}
    \mathcal{F}_i (\mathcal{B})  = \sum^{k_p}_{j=1} |\braket{e_i|\phi_j}|^2  ;\;\;\; i=1,2, \dots N
\end{equation*}

In the \cref{fig:overlap_basis_noise}, we plot the overlap of basis with all the eigenstates of the Hamiltonian system described above. 
Basis formed using truncated ASP is able to span the low-energy subspace, showing high overlap with the low-energy states 
whereas, the Krylov basis formed using powers of the operator $H$ randomly spans states. 
As the amount of noise is increased, overlap with other excited states increases for EC basis, 
and thus condition number of the basis decreases with noise.
For the Krylov basis, however, noise introduces other states at random, and, the condition number remains comparable or increases. 

Using the Krylov basis, we aim for the basis to span the ground state at the desired target Hamiltonian, and,
while using EC with truncated state prep methods, we aim to span the ground states across the parameter spectrum.
As a result, EC basis has high overlap with multiple lowest energy states at different parameter points and other low energy states, whereas, 
the Krylov basis has contributions from all the states. 

\subsection{Measurement error}

To account for the measurement errors in the noise analysis, we introduce an error to all the matrix element values. 
Measurement errors could include two main types of errors including a readout error caused by an error in reading the output and another due to $T1$ decoherence error causing bit-flip from $0 \rightarrow 1$. 
As mentioned later in the discussion, there could be multiple ways of calculating the matrix elements on a quantum device. 
For our measurement error analysis, we considered the Hadamard test protocol. 
To do so, we consider all the samples generated by implementing the noise model and change the expected matrix element values according to the following equation.
We consider the values of readout errors from a ref \cite{funcke2022measurement} as $p_{0 \rightarrow 1} = 0.01$ and $p_{1 \rightarrow 0} = 0.03$.  
To account for $T_1$ error decoherence, we add a max $p_{T1} = 0.03$ that cause a biased bit-flip from $\ket{1} \rightarrow \ket{0}$. 

\begin{figure}[htpb]
    \centering
    \includegraphics[width = 0.48\textwidth]{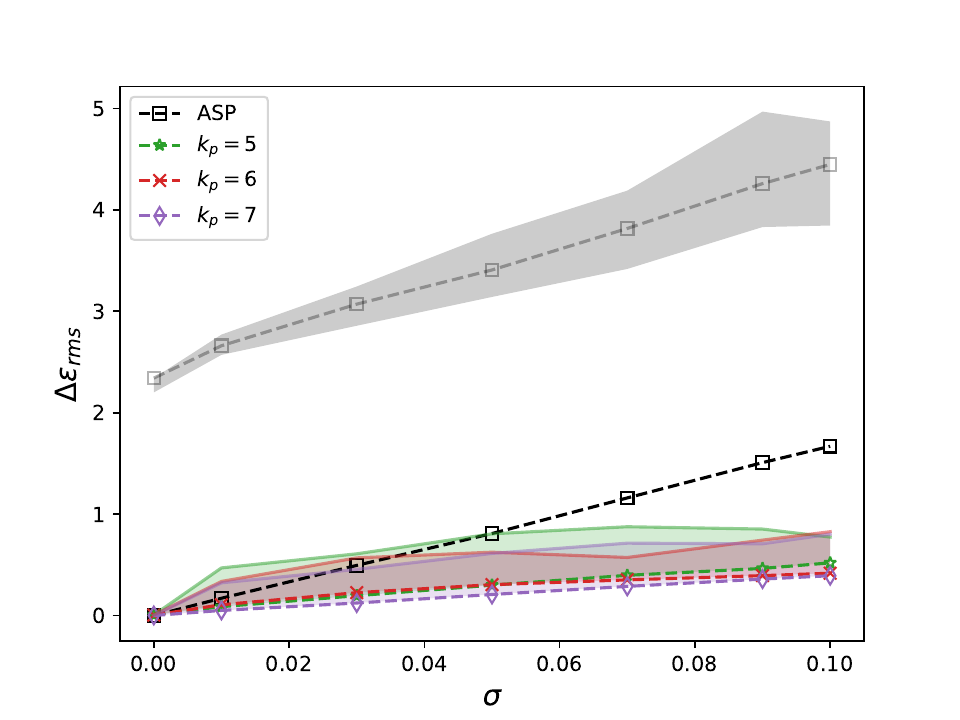}
    \caption{Noise analysis with measurement error. The shaded region shows the change in error including the measurement error. }
    \label{fig:noise_with_me}
\end{figure}

For comparison with ASP, we included measurement errors in measuring Hamiltonian expectation values for a state produced by an ASP circuit by introducing a confusion matrix as explored by Rigetti in \cite{forestbenchmarking}.
In the \ref{fig:noise_with_me}, the shaded region shows values with measurement error for ASP and with EC.

\end{document}